\documentclass[aps,prd,twocolumn,prd,showpacs,showkeys,preprintnumbers,bibnotes,floatfix,longbibliography,notitlepage,nofootinbib,superscriptaddress,superscriptgaddress,amsmath
]{revtex4-2}

\def\apj{ApJ}%
\def\apjl{ApJ}%
\def\apjs{ApJS}%
\def\aap{A\&A}%
\def\jcap{JCAP}%
\def\mnras{MNRAS}%
\def\prd{Phys.\ Rev.\ D}%
\def\prl{Phys.\ Rev.\ Lett.}%
\def\pasj{PASJ}%

\usepackage{graphicx}
\usepackage{dcolumn}
\usepackage{bm}
\usepackage{tabularx}

\usepackage[colorlinks]{hyperref}
\hypersetup{
    pdfnewwindow=true,      
    colorlinks=true,       
    linkcolor=blue,          
    citecolor=blue,        
    filecolor=blue,      
    urlcolor=blue        
}

\usepackage{booktabs}
\usepackage{xcolor}
\usepackage{listings}
\usepackage{calligra}
\usepackage{xspace}

\DeclareMathAlphabet{\mathcalligra}{T1}{calligra}{m}{n}
\DeclareFontShape{T1}{calligra}{m}{n}{<->s*[2.2]callig15}{}

\newcommand{\er}{E_R}
\newcommand{\ud}{\mathrm{d}}
\newcommand{\mx}{m_\chi}

\newcommand{\sigsi}{\sigma_{\text{\tiny{SI}}}}
\newcommand{\gaia}{\textit{Gaia}\xspace}
\newcommand{\galpy}{\texttt{galpy}\xspace}

\begin{document}


\title{Dark matter limits from the tip of the red giant branch of globular clusters}

\author{Haozhi Hong}
\email{xili@phas.ubc.ca}
\affiliation{
Department of Physics and Astronomy, University of British Columbia,\\
Vancouver, British Columbia V6T 1Z1, Canada
}
\author{Aaron C. Vincent}
\email{aaron.vincent@queensu.ca}
\affiliation{
Department of Physics, Engineering Physics and Astronomy,\\
Queen’s University, Kingston ON K7L 3N6, Canada
}
\affiliation{%
Arthur B. McDonald Canadian Astroparticle Physics Research Institute, Kingston ON K7L 3N6, Canada
}
\affiliation{Perimeter Institute for Theoretical Physics, Waterloo ON N2L 2Y5, Canada}


\begin{abstract}
Capture and annihilation of WIMP-like dark matter in red giant stars can lead to faster-than-expected ignition of the helium core, and thus a lower tip of the red giant branch (TRGB) luminosity. We use \gaia data to place constraints on the dark matter-nucleon cross section using 22 globular clusters with measured TRGB luminosities, and place projections on the sensitivity resulting from 161 clusters with full phase space distributions observed by \gaia. Although limits remain weaker than those from Earth-based direct detection experiments, they represent a constraint that is fully independent of dark matter properties in the Solar neighbourhood, probing its properties across the entire Milky Way galaxy. Based on our findings, it is likely that the use of the TRGB as a standard candle in $H_0$ measurements is very robust against the effects of dark matter.

\end{abstract}

\maketitle

\section{Introduction}
In the $\Lambda$CDM framework, observations across galactic scales and beyond indicate that
approximately 84\% of mass density in our universe is in the form of cold, collisionless dark matter \cite{Planck:2018vyg}. The prototypical dark matter candidate is the weakly interacting massive particle (WIMP). The classical WIMP has a GeV-scale or higher mass, and would have been produced from thermal freeze-out in the Early Universe, requiring an annihilation cross section $\langle \sigma v \rangle \sim 10^{-26}$ cm$^{3}$s$^{-1}$, approximately independent of mass. Two-to-two annihilation into Standard Model  (SM) particles implies WIMP-nucleon elastic scattering to exist in most models, though the scattering cross section $\sigma_{\chi}$ can vary by orders of magnitude depending on the microphysics of the interaction. 

Thus far, direct searches for dark matter-nucleon have yielded null results, pushing limits on spin-independent scattering downward of $\sigma_\chi < 10^{-47}$ cm$^{2}$ between $m_\chi = 20$ and 40 GeV, \cite{LZ:2022ufs}. Limits become weaker at high mass due to limited exposure, and at low mass from kinematical limitations. Searches for elastic scattering of dark matter can be augmented by solar capture and subsequent annihilation to neutrinos, detectable by Earth-based  observatories such as IceCube~\cite{IceCube:2012ugg,IceCube:2012fvn,IceCube:2016yoy}, ANTARES~\cite{ANTARES:2016xuh} and SuperKamiokande \cite{Super-Kamiokande:2015xms}. Nonetheless, it is germane to ask whether searches may be performed in a way that is entirely independent of Earth-based systematics, such as the local (in space and time) density and velocity distribution of dark matter near the solar system.

Stellar populations have long been used to infer properties and place limits on models of new physics \cite{Raffelt:1996wa}. Here, we focus on one particular scenario. 
Gravitational capture of dark matter in stars due to elastic scattering below the local escape velocity inevitably leads to accumulation of large quantities over the long lifetime of stars \cite{Press1985CaptureParticles,1987ApJ...321..571G}. In sufficient quantities, dark matter can transport heat \cite{Spergel1985EffectInterior,Nauenberg1987,Gould1990}, leading to structural rearrangement. By modifying the thermal profile near the hydrogen-burning core, slight changes in the main sequence turnoff can occur \cite{Lopes:2019jca}, and the main sequence lifetime can be significantly reduced \cite{Raen:2020qvn}.

If the dark matter $\chi$ self-annihilates, far fewer particles can be accumulated, but the effects on stellar energetics can be even more drastic, injecting 2$m_\chi$ of heat per annihilation  \cite{SalatiSilk89, BouquetSalati89a, Moskalenko07, Bertone07, Spolyar08, Fairbairn08, Scott08a,Scott:2008ns, Iocco08a, Iocco08b, Casanellas09, Ripamonti10, Zackrisson10a, Zackrisson10b, Scott11,Turck-Chieze:2012zie,John:2024thz}, leading to changes in the core structure. 

Indeed, Ref. \cite{Lopes:2021jcy} has shown that for a large enough capture rate of dark matter in red giant stars, the additional injected heat from annihilation can lead to premature ignition of helium burning, thus measurably lowering the \textit{tip of the red giant branch} (TRGB) luminosity on the Hertzprung-Russel (temperature-luminosity) diagram (Ref.~\cite{Dessert:2021wjx} also placed constraints on macroscopic  $m > 10^{17}$ g dark matter from heating the He core with its kinetic energy). 

As main sequence stars exhaust their available hydrogen fuel in the core, the inert helium core collapses leading to gravothermal heating, a higher luminosity, but lower envelope temperature as the outer layers of the star expand. As these stars climb the red giant branch, their luminosity becomes nearly entirely independent of the total stellar mass, and instead is governed by the mass of the helium core around which hydrogen shell fusion takes place, further increasing the core mass and precipitating more collapse. Once the core temperature reaches $\sim 10^8$K, helium fusion ignites, halting the star's climb. This happens when the core mass reaches $\sim 0.48$ M$_\odot$, with a weak dependence on metallicity but almost entirely independent of other stellar parameters, meaning that the TRGB luminosity is a \textit{standard candle} (see e.g. \cite{kippenhahn1990stellar} for more details). Because of this, the TRGB has for example been used on the distance ladder required for measurements of the Hubble parameter $H_0$ and the accelerated expansion of the Universe \cite{Freedman2019}. 

Because the effect of DM annihilation on the TRGB strongly depends on the local density of dark matter, we seek to compare the TRGB in different environments to search for any correlation between the dark matter ``bath'' a given population of stars experiences, and its TRGB luminosity. For this task, globular clusters (GCs) are ideal laboratories. These old, compact collections of upward of 10$^6$ stars exist on a variety of orbits around our galaxy. While some have wide, sweeping orbits, others exist on trajectories that have brought them close to the galactic centre and its large density of dark matter particles. With enough information about the GC's position and current velocity, and the Milky Way's gravitational potential, we may reconstruct a likely trajectory and thus infer roughly how much of the dark matter halo it has passed through, and what effects this should have had on the stars in each of these clusters.

Recent measurements revealed in \textit{Gaia}'s DR3 \cite{brown2021gaia,2023A&A...674A...1G} have provided high precision distance and velocity determinations on 161 GCs, reported in Ref. \cite{Vasiliev_2021}. Combined with recent estimates of the Milky Way gravitational potential, we use these measurements as ``initial conditions'', and evolve their orbits back in time. We may then include this additional heat from dark matter capture and annihilation into MESA simulations of sub-solar mass stars, leading to a predicted TRGB for each GC trajectory, as a function of the dark matter mass and cross section. These can finally be compared with measured TRGB luminosities. We will set limits based on the measured TRGB magnitudes of a subset of 22 of these GCs reported in Ref. \cite{Straniero:2020iyi}, who used photometric measurements from the Hubble Space telescope and ground-based observatories, with \textit{Gaia}-calibrated distances. We will also show sensitivity curves assuming the full \textit{Gaia} sample could be measured, and under the assumption that the TRGB could be measured more precisely in the future. 

Because the sampled dark matter density varies by many orders of magnitude, the inferred limits will turn out to be quite insensitive to the details, or errors, on the modelling of the integrated orbits. We will find that the choice of dark matter halo does make a large difference, with a contracted halo leading to larger sampled densities, and therefore stronger limits than a classic NFW. 

This article is structured as follows: we first describe our parametrization of the Milky Way gravitational potential and dark matter distribution. In Sec. \ref{sec:capture} we describe the capture of dark matter in GC red giants, as well as our numerical stellar simulations. Our results are presented in Sec.~\ref{sec:results}, and we conclude in Sec.~\ref{sec:conclusion}.

\section{The Milky Way gravitational potential}
\label{sec:potential}
The amount of dark matter captured in a star (and therefore, injected as heat in the core) is directly proportional to the local dark matter density $\rho_\chi(r)$, where $r$ is the distance to the galactic centre. Depending on its orbital history, each globular cluster will sample a different amount of dark matter. To assess each cluster's ``dark matter history'', we take their current positions and velocities as measured using \gaia data \cite{brown2021gaia}, and integrate their equations of motion backward in time over a few Gyr, to represent the approximate lifetime on the red giant branch. To do so, we require a detailed model of the Milky Way gravitational potential, including the baryonic (stars, gas) and dark matter components. The latter contributes both to the potential, and to the locally-sampled dark matter $\rho_\chi (r)$. 

A number of recent works have used kinematic measurements to produce a parametrization of the Milky Way mass distribution. We will present our results in terms of two benchmark models: De Salas et al.'s \cite{2019JCAP...10..037D} B1+NFW, and the contracted halo model described in Cautun et al. \cite{Cautun:2019eaf}. These respectively represent a ``traditional'' Navarro-Frenk-White (NFW) dark matter distribution, and a dark matter-plus-baryon simulation-motivated halo that has a steeper central slope.  At the end of this section, we will further discuss the possible time-dependence of the potential.

de Salas et al. \cite{2019JCAP...10..037D} model the potential $\Phi(\vec r)$ using a bulge, a thin disk, a thick disk, and an NFW dark matter halo. They use rotation curve data from \gaia DR2 \cite{2019ApJ...871..120E} and perform an MCMC parameter space search over a number of different models. Here, we  employ the B1+NFW model. For B1, the bulge is modeled as a Plummer profile. In cylindrical coordinates $(R,z)$,
\begin{equation}
    \Phi_{\rm Plummer} = -\frac{GM_{\rm bulge}}{\sqrt{R^2+z^2+R_b^2}},
\end{equation}
where $M_{\rm bulge}$ is the bulge mass $R_b$ is a constant describing its radial extent. Both stellar disks use a Miyamoto-Nagai potential \cite{1975PASJ...27..533M}: 
\begin{equation}
    \Phi_{\rm MN}(r,z) = -\frac{GM_{\rm disk}}{\sqrt{R^2+\left( R_d+\sqrt{z^2+z_d^2} \right)^2}} ,
\end{equation}
with a distinct disk mass value $M_{\rm disk}$ for each disk component, along with a radial and height scales, $R_d$ and $z_d$. The dark matter is modeled as an NFW profile:
\begin{equation}
    \rho_{\rm NFW}(r) = \rho_0\left(\frac{r}{r_s}\right)^{-1}\left(1+\frac{r}{r_s}\right)^{-2},
    \label{eq:nfw}
\end{equation}
where $r_s$ is the scale radius, and $\rho_0$ is a normalization. The best fit value of each parameter can be inferred from Table I of \cite{2019JCAP...10..037D}, along with their $1\sigma$ confidence limits, which we will use when examining the impact of error propagation on the dark matter history of each cluster, and thus on the limits that we set.

Cautun et al. \cite{Cautun:2019eaf} perform a similar analysis of the MW rotation curve, also using \gaia DR2, modeling a bulge, a thin disk, a thick disk, and additionally include an atomic hydrogen disk and a molecular gas disk, as well as a circumgalactic medium (CGM) component. They examine two scenarios: one with a pure NFW dark matter profile, and the other with a ``contracted'' halo, as motivated by the Auriga \cite{2017MNRAS.467..179G}, APOSTLE \cite{2016MNRAS.457..844F,2016MNRAS.457.1931S} and EAGLE \cite{2015MNRAS.446..521S} suites of Milky Way analog simulations. The contracted halo is modeled as an NFW profile matching dark matter-only simulations, rescaled at each radial point based on the results of their hydrodynamical simulations.

The Cautun bulge is a McMillan profile \cite{2017MNRAS.465...76M}:
\begin{equation}
    \rho_{\text{bulge}} = \frac{\rho_{0,\text{bulge}}}{(1 + r'/r_0)^\alpha} \exp\left[-\left(\frac{r'}{r_{\text{cut}}}\right)^2\right],
\end{equation}
with $r' = \sqrt{R^2 + \left({z}/{q}\right)^2}.$ The stellar disks are exponentials:
\begin{equation}
    \rho_\mathrm{disk}(R, z) = \frac{\Sigma_0}{2 z_d} \exp\left(-\left|\frac{z}{z_d}\right| - \frac{R}{R_d}\right),
\end{equation}
and the gas disks are doughnut-shaped:
\begin{equation}
    \rho_\mathrm{gas}(R, z) = \frac{\Sigma_0}{4 z_d} \exp\left(-\frac{R_m}{R} - \frac{R}{R_d}\right) \operatorname{sech}^2\left(\frac{z}{2 z_d}\right).
\end{equation}
Finally, the CGM is modeled as a power law $\rho_{\rm CGM} \propto r^{-1.46}$. The constant parameters $r_0, q, r_\mathrm{cut}, z_d, R_d, R_m$ that govern the shapes of each of these distributions can be found in Table 1 of Ref.~\cite{Cautun:2019eaf}. The best fit normalization of each component can be found in Table 2 of the same reference.

The choice of de Salas B1+NFW and the contracted Cautun model is meant as a means to discern the effect of the dark matter halo shape on our final results. Cautun et al.'s analysis examined a ``pure'' NFW  (Eq. \ref{eq:nfw}) in addition to the contracted halo, finding that do not give significantly different fits to the rotation curve data. In contrast, we have found that using the Cautun et al. pure NFW model will result in weaker limits on the dark matter microphysics that are very close to those derived with B1+NFW, suggesting that the choice of baryonic profile is far less important than the dark matter parametrization. We will furthermore show the effects of error propagation for NFW, but not the contracted halo, since the latter presents numerical challenges. 

To implement each of these models, we use the \galpy\footnote{\url{http://github.com/jobovy/galpy}} software suite \cite{2015ApJS..216...29B}. We will use \galpy to implement these potential functions, and to integrate the dynamical trajectories, with the exception of the time-dependent potential which we describe near the end of the next subsection.

\subsection{Globular cluster trajectories}
For GC trajectory integration, we use the processed Gaia data provided by \cite{Vasiliev_2021}. The processed Gaia data contain the kinematic properties of 170 GCs,  161 of which have full six phase-space kinematic parameters. \galpy includes functions to integrate orbits from  user-specified initial conditions and gravitational potentials sourced from dark matter and baryonic matter as detailed above. 
From these orbits, we can extract each GC's local dark matter density $\rho_\chi(t)$, the speed $v_\star(t)$ and the local DM velocity distribution dispersion $v_0(t)$ for each GC as a function of time. We calculate $v_0$ as a function of position using the spherical Jeans equation \cite{binney2011galactic}.

\begin{figure}
   \hspace{-1.cm} \includegraphics[width=0.52\textwidth]{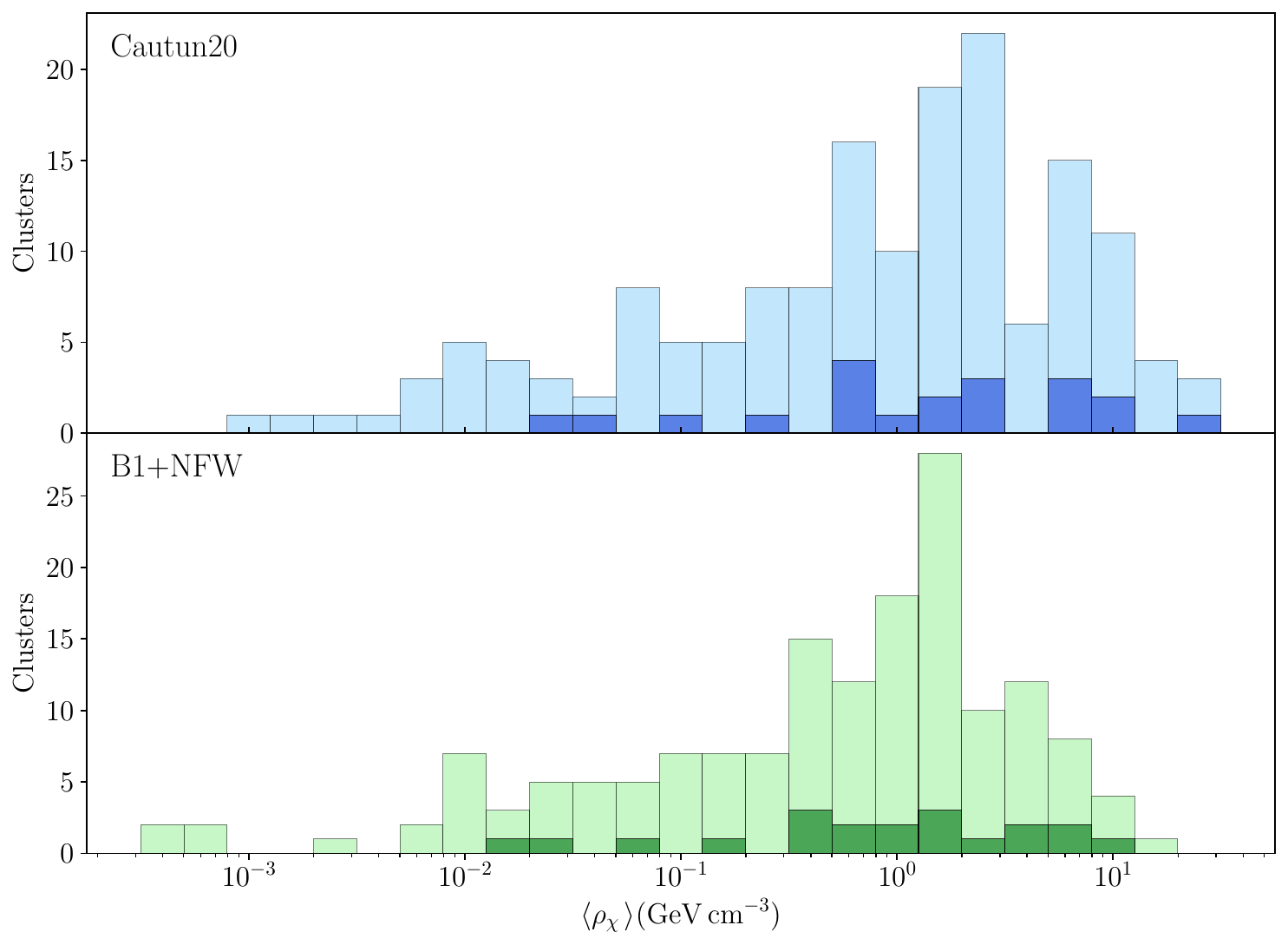}
    \caption{ The dark matter density encountered by Milky Way globular clusters in their time-averaged orbital history, for all 161 clusters with 6d phase space data from \gaia (light colors), and for the 22 clusters with TRGB magnitudes \cite{Straniero:2020iyi} that we use to set the limits (dark colors). Top panel uses the Cautun et al. \cite{Cautun:2019eaf} contracted halo model; bottom panel is with the de Salas et al. \cite{2019JCAP...10..037D} B1+NFW model.}
    \label{fig:histogram}
\end{figure}

 Fig. \ref{fig:histogram} is a histogram of the time-average dark matter density $\langle \rho_\chi \rangle$ as encountered by each globular cluster, for the Cautun contracted halo (top), and the de Salas et al. B1+NFW model (bottom). The smaller, darker distributions show the clusters for which we have TRGB luminosity measurements \cite{Straniero:2020iyi}, while the lighter distributions show the full set of 161 GCs. The crucial feature of this plot is that the distribution spans many orders of magnitude in $\rho_\chi$. This fact makes limit-setting fairly robust against any systematic uncertainty. As mentioned previously, the choice of halo parametrization does make a significant difference. The Cautun contracted halo displays an overall larger DM density, and will therefore lead to a higher capture rate for a given cross section, and thus to stronger limits. This is not due to differences in baryonic modeling or total mass; rather, with a fixed local density $\rho(R_\odot) \sim 0.3$~GeV~cm$^{-3}$ the contracted profile is steeper than $r^{-1}$ in the central regions, and falls off less quickly than NFW at large radii, such that more dark matter is sampled overall. This explanation is borne out by Cautun's pure NFW profile, which yields very similar results to de Salas B1+NFW. 

Given that stellar evolution to the TRGB takes place over Gyr timescales, the time dependence of the dark matter profile could in principle prove important. To test this, we also integrate NFW profiles modified with a time-dependent concentration parameter $c(t)$, following the results obtained in Dutton et al.~\cite{Dutton:2014xda}. Based on cosmological simulations, and keeping the baryonic matter distributions from de Salas. Because the Dutton et al. dark matter profile is not calibrated to match the present-day Milky Way rotation curves, the densities at $t = 0$ (today) are not expected to match the other models. Rather, this should serve as a way to test the effect of a time-dependent NFW halo.

\begin{figure}
    \centering
    \hspace{-1.cm} 
    \includegraphics[width=0.53\textwidth]{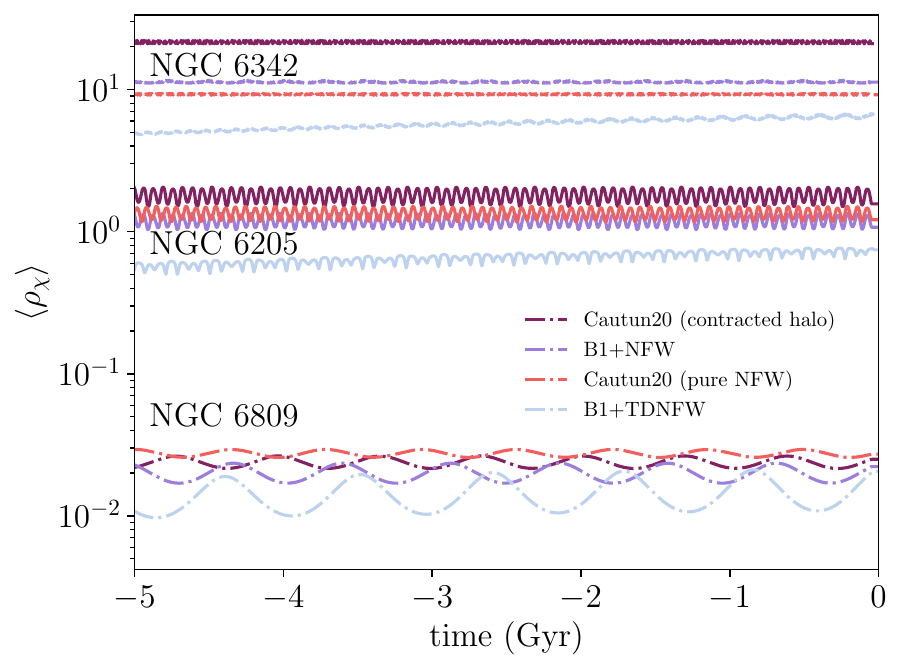}
    \caption{65 Myr Moving average of the dark matter density encountered by three different globular clusters that contribute to the limits presented here, under different halo model assumptions. The three clusters are chosen to represent a high, medium and low exposure to dark matter. Limits presented here are based on the Cautun20 contracted halo model, and the de Salas B1+NFW models. For comparison, we also show the Cautun pure NFW model (which gives very similar results to B1+NFW), and a time-dependent halo model. The latter is not normalized to today's local density, but serves to illustrate that the time-dependent halo contraction does not markedly affect $\langle \rho_\chi \rangle$. }
    \label{fig:rho_vs_time}
\end{figure}

Fig.~\ref{fig:rho_vs_time} shows the dark matter density sampled by three clusters from the sample examined here, NGC 6342, NGC 6205 and NGC 6809, respectively representing clusters that encounter a high, intermediate, and low dark matter density. We compare the moving average (with a 65 Myr window) of the local dark matter density for the static Cautun20 contracted halo, the de Salas B1+NFW, as well as Cautun's pure NFW and the time-dependent NFW profile labeled B1+TDNFW. We draw three main conclusions from Fig. \ref{fig:rho_vs_time}: 1) the separation between dark matter densities encountered by each cluster remains well-defined from model to model; 2) the Cautun et al. pure NFW and the B1+NFW result in very similar capture rates, and 3) the time-dependence of the halo concentration only results in an O(1) change in density over the 5 Gyr range plotted here. 

\section{Stellar capture of dark matter}
\label{sec:capture}
We consider dark matter particles $\chi$ which scatter elastically with nucleons via a constant, spin-independent cross section $\sigsi$. 
The capture rate of dark matter in stars was described by Gould \cite{Gould87a}:
\begin{equation}
C_\star(t) = 4\pi \int_0^{R_\star} r^2 \int_0^\infty \frac{f_\star(u)}{u} w \Omega(w) \ud u\ud r,
\label{caprate}
\end{equation}
where $f_\star(u)$ is the dark matter velocity distribution far from the star, $w(r) \equiv \sqrt{u^2 + v_e(r)^2}$ is the velocity of dark matter that has fallen to a distance $r$ from the centre of the star, boosted by the local escape velocity $v_e(r) = \sqrt{-2\phi(r)}$, where $\phi(r)$ is the gravitational potential. $\Omega(w)$ is the rate of scattering from velocity $w$ to any velocity below $v_e(r)$.  
\begin{align}
 \Omega(w) =& \frac{2}{m_\chi w} \sum_i \sigma_{N,i} n_i(r,t) \frac{\mu_{i,+}^2}{\mu_i}\Theta\left(\frac{\mu_i v_{\rm esc}^2}{\mu_{i,-}^2} - u^2 \right) \nonumber \\ \times& \int_{\mx u^2/2}^{\mx w^2 \mu_i/2\mu_{i,+}^2} |F_i(\er)|^2 \ud \er,
 \label{eq:omega}
\end{align} 
where the sum is over nuclear species $i$. $\sigma_{N,i} = \sigsi A_i^2 (m_i/m_p)^2(m_\chi+m_p)^2/(m_\chi + m_i)^2$ is the dark matter-nucleus elastic scattering cross section, where $m_p$ is the proton mass, $m_i$ is the mass of the nucleus, and $A_i$ is the atomic mass number. We use the usual Helm form factor
\begin{equation}
|F_i(\er)|^2 = \exp{\left(-\frac{\er}{E_i}\right)},
\end{equation}
with $E_i = 3/(2m_i \Lambda_i)$ and $\Lambda_i = [0.91(m_i/\mathrm{GeV})^{1/3} + 0.3]$ fm. 
The velocity distribution of Milky Way halo dark matter in the frame of the star is:
\begin{equation}
f_\star(u) = \left(\frac32\right)^{3/2}\frac{4\rho_\chi u^2}{\pi^{1/2}m_\chi u_0^3}e^{-\frac{3 (u_\star^2+u^2)}{2u_0^2} } \frac{\sinh (3 u u_\star/u_0^2)}{3 u u_\star /u_0^2}. \label{eq:fofu}
\end{equation}
We will not consider the proper motions of stars within each GC. Some claims have further been made about large dark matter densities inside globular clusters, however, these are not robust \cite{Moore:1995pb,Saitoh:2005tt}. We therefore do not include any intrinsic dark matter to each globular cluster, or any overdensity from the clusters' gravitational pull. These effects could in principle strengthen the bounds presented here. 

Since $f_\star(u) \propto \rho_\chi$, the local dark matter density, and $\Omega(w) \propto \sigsi$, then $C_\star(t)$ should na\"ively scale as $\sigsi \rho_\chi$. However, at large values of $\sigsi$, the capture rate will begin to saturate as all available dark matter at some radius $r$ has already been captured at higher radii. This effect should eventually plateau at the \textit{geometric limit}, where every dark matter particle that intersects the star is captured, and further increasing $\sigsi$ thus has no effect.

Several different methods have been considered to account for saturation. A simple cap on the capture rate at $C = C_\mathrm{max}$ overestimates capture as it approaches this limit. For large cross sections, a multiple-scatter formalism \cite{Bramante:2017xlb,Leane:2023woh,Ilie:2024sos} is often employed, with the simplifying assumption that energy loss after each scattering event is approximately equal to the average. This formalism works well in dense, uniform objects such as white dwarfs or even planets (though some issues have been raised \cite{Bell:2024qmj}). Red giants, however, contain a very diffuse atmosphere but a high-density core. 

We instead use the approach of \cite{Gould92OpticalDepth,Busoni:2017mhe}, which is to suppress the incoming dark matter flux at each radius $r$ by an optical depth factor:
\begin{equation}
    \eta(r) =\frac{1}{2}\int_{-1}^1 dz e^{-\tau(r,z)}\label{eq:extinction},
\end{equation}
where $\tau(r,z)$ is the optical depth to the surface at height $r$, in the line-of-sight direction parametrized by the angle $\beta$ from the radial vector:\footnote{See Fig. 4 of \cite{Busoni:2017mhe} for an illustration of the geometry.}
\begin{equation}
    \tau(r,z)=\int_{rz}^{\sqrt{R^2-r^2(1-z^2)}} dx \sum_i n_i(r') \langle\sigma_{i,Tot}\rangle\label{eq:opticdepth},
\end{equation}
with  $z \equiv \cos\beta$ and $r'^2 \equiv x^2+r^2(1-z^2)$. The sum here is over species, and the thermal average of the total cross section is $2 \sigsi$. 

However, including the factor of $\eta(r)$ in the integrand of Eq.~\eqref{caprate} will smoothly suppress the capture rate, but it will asymptote at the \textit{single-particle saturation limit}, rather than the geometric limit. Since the suppression in Eqs. (\ref{eq:extinction}-\ref{eq:opticdepth}) removes all particles that have scattered from the flux, regardless of whether that scattering event led to capture, it does not account for particles that scatter without being captured falling to deeper (and denser) radii and possibly being captured there. Eq. \eqref{eq:opticdepth} works well for stars such as the Sun (as shown in Ref. \cite{Busoni:2017mhe}), but in the large, diffuse atmosphere of a red giant, the single-particle saturation rate is much lower than the geometric limit. To address this, we correct the optical depth calculation to include only scattering that leads to capture in the averaged cross section:
\begin{equation}
    \tau(r,z)=\int_{rz}^{\sqrt{R^2-r^2(1-z^2)}} dx  \int du \Omega(w)\frac{wf_\star(u)}{u}\label{eq:opticdepthcorr}.
\end{equation}
This will lead to a total saturation that is closer to the geometric limit.\footnote{We note a similar approach was taken in the recent Ref. \cite{Bell:2024qmj}.} However, it does very slightly ``overcompensate'', since it does not take into account the modification of the velocity distribution $f_\star(u)$ due to scattering without capture, or equivalently, does not remove the flux of particles at radius $r$ that would have been captured by multiple interactions. This overcompensation is stronger at low dark matter mass, and becomes O(1) around $m_\chi = 10$ MeV. For this reason, we will restrict ourselves to $m_\chi > 100$ MeV.
\begin{figure}
    \centering
    \includegraphics[width=0.5\textwidth]{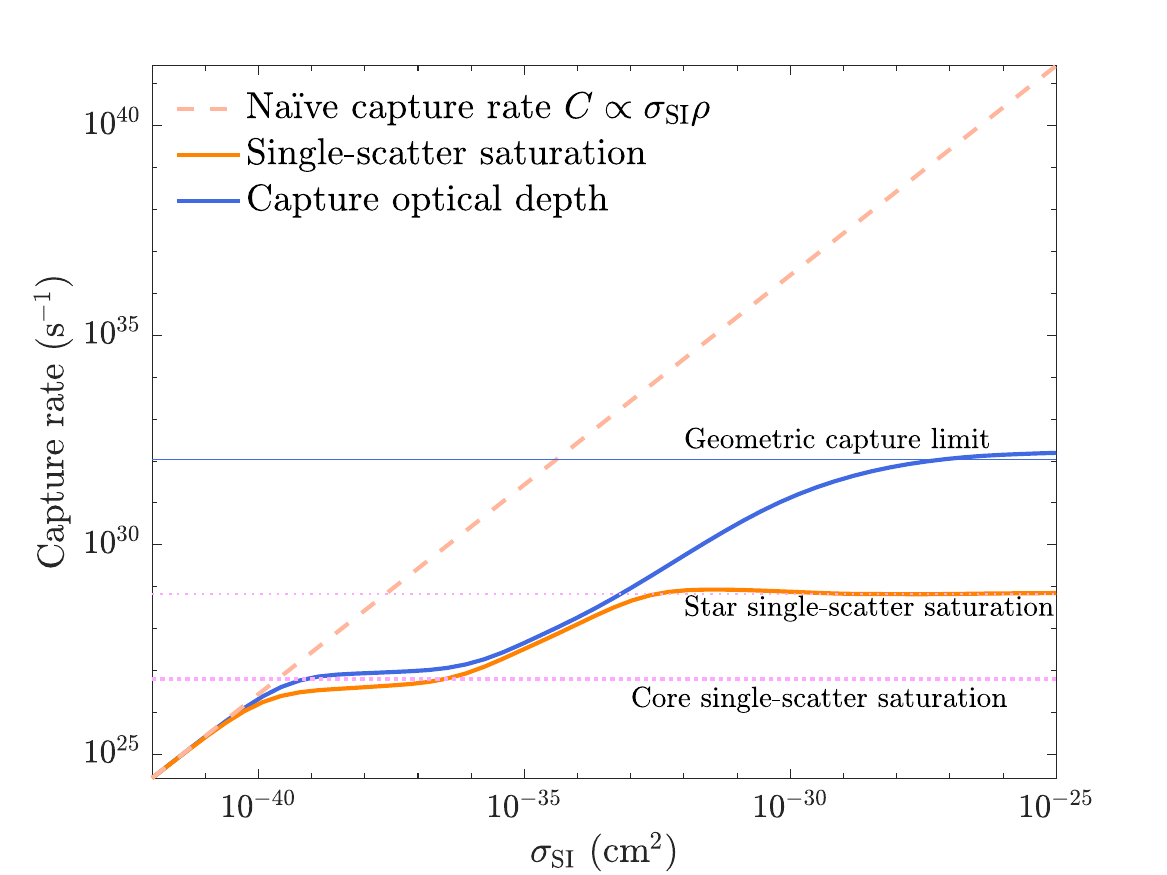}
    \caption{Capture rate of dark matter on a 0.8 $M_\odot$ red giant at the local dark matter density $\rho_0 = 0.4$ GeV cm$^{-3}$, for $m_\chi = 20$ GeV. Dashed diagonal line: na\"ive capture rate \eqref{caprate} without taking saturation into account; orange line: capture rate assuming only single-scatter capture \eqref{eq:opticdepth}; blue line: full capture rate using the adjusted optical depth suppression \eqref{eq:opticdepthcorr}, which correctly converges to the geometric capture limit (all dark matter intersecting the stellar disk is captured).}
    \label{fig:saturation}
\end{figure}

Fig. \ref{fig:saturation} summarises this discussion, showing the capture rate of a $m_\chi = 20$ GeV WIMP on a 0.8 $M_\odot$ red giant assuming local (solar) dark matter halo values. Other parameters yield qualitatively equivalent results. The light dashed line shows the ``na\"ive'' capture rate, i.e. using Eq. \eqref{caprate} with no saturation included. The orange line uses the result of Ref. \cite{Busoni:2017mhe} and the full optical depth suppression \eqref{eq:opticdepth}, showing two saturation plateaus: one on the dense helium core, and a second for the full star. This saturation limit (also calculated analytically, pink dotted line) lies well below the geometric saturation rate. Finally, the blue curve uses the Eq. \eqref{eq:opticdepthcorr} to compute the optical depth, correctly saturating at the geometric limit. We use this parametrization in the remainder of this work. 

We will search for effects of dark matter capture and annihilation for masses between $m_\chi = 0.1$ and $10^4$ GeV. Estimates of the evaporation mass using $E_c/T_c \simeq 30$ \cite{Garani:2021feo}, where $E_c$ and $T_c$ are respectively the escape energy and temperature at the centre of the star, yields $m_{\mathrm{evap}} \simeq 1$ GeV for a 0.8 $M_\odot$ red giant. However, we expect this to be somewhat lower, as the optical depth to the surface $\tau(r = 0)$ for cross sections considered here is significantly larger than one.

\subsection{Modeling capture in globular cluster red giants}

In order to model the effects of DM annihilation on the TRGB, we employ the MESA \cite{2011ApJS..192....3P,2013ApJS..208....4P,2015ApJS..220...15P,2018ApJS..234...34P,2019ApJS..243...10P,2023ApJS..265...15J} module developed and publicly released as part of Ref. \cite{Lopes:2021jcy}, which we have modified to run on MESA r23.05.1, and add the ability to run simulations with a time-dependent stellar velocity, local DM density and local dark matter dispersion velocity. This module computes the capture rate of dark matter at every time step, and injects extra energy into the stellar plasma assuming capture and annihilation have equilibrated, a reasonable assumption given the large capture rates considered here. Going forward we set the fraction $f_\chi$ of the total annihilation energy $2m_\chi$ going into heating the plasma to $f_\chi = 1$. If annihilation to neutrinos or other dark sector particles were to occur this fraction would lower, in turn weakening our bounds by $f_\chi^{-1}$. Our results are close to those presented in \cite{Lopes:2021jcy}, although we find a slight but systematic smaller effect using MESA r23.05.1 in comparison with their results.

\begin{figure}[h]
    \includegraphics[width=0.54\textwidth]{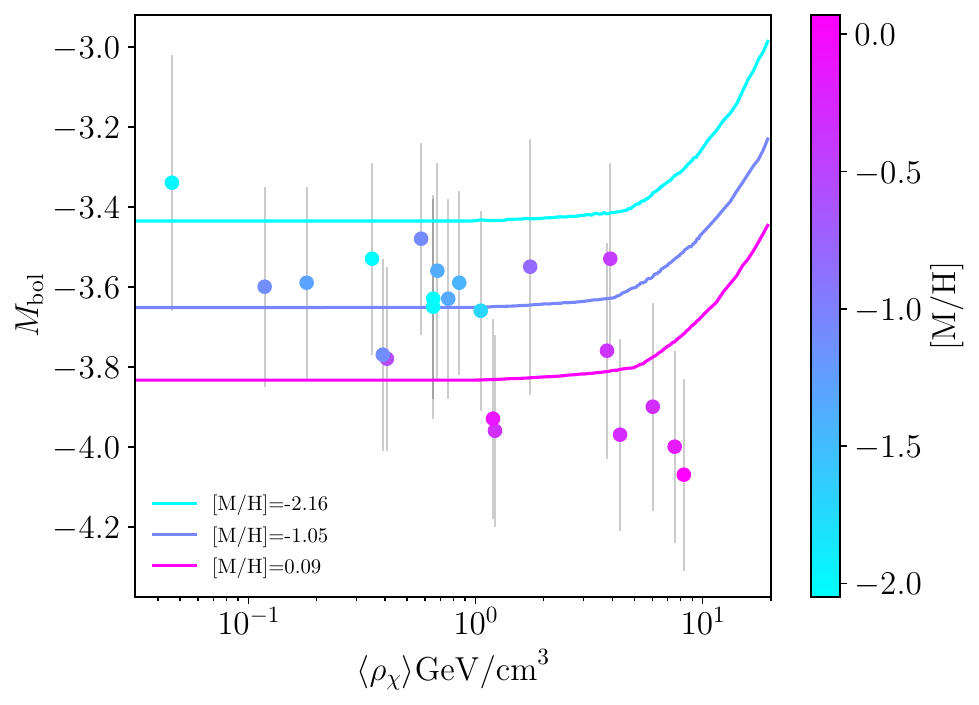}
    \caption{Lines: the impact of dark matter capture and annihilation on the bolometric magnitude of the tip of the red giant branch, as a function of the average dark matter encountered by a red giant star, for dark matter mass $m_\chi = 10$ GeV, and $\sigsi = 10^{-28}$ cm$^2$, the 90\% CL limit for this mass. The three lines show three different metallicity values. Points: the $\langle \rho_\chi \rangle$---$M_{\mathrm{bol}}$ values for the 22 globular clusters \cite{Straniero:2020iyi} used in this analysis.}
        \label{fig:Mbolvsrho}
\end{figure}

The solid lines in Fig. \ref{fig:Mbolvsrho} shows the effect of dark matter on the TRGB magnitude $M_\mathrm{bol}$
\begin{equation}
    M_\mathrm{bol} \equiv -2.5\log_{10}(L/L_\odot) + 4.74,
\end{equation}
where $L$ is the TRGB luminosity, for three different values of the metallicity [M/H], for a dark matter mass $m_\chi = 10$ GeV, and a cross section $\sigsi = 10^{-27}$ cm$^2$, corresponding to the 90\% CL limit for this mass. The points in Fig. 4 are the 22 GCs from Ref.~\cite{Straniero:2020iyi} used to set the limits in this article, with their corresponding metallicities given by the color scale. 

Strictly speaking, the capture rate in each cluster star depends on the integrated trajectory of the cluster's position, velocity with respect to the frame of the galaxy, the star's individual velocity within the cluster, and the local dark matter dispersion velocity. Since dark matter affects the TRGB age of the star, a trial and error approach with multiple simulations is needed because stellar evolution is modeled forward in time, while trajectories are integrated backward from the current conditions. Since many MESA simulations are necessary for every individual cluster and for every dark matter mass and cross section, it would be computationally infeasible to also include the time-dependent capture rate in every simulation that we use to compute limits.

\begin{figure}
     \centering 
     \hspace{-.7cm}
\includegraphics[width=0.51\textwidth]{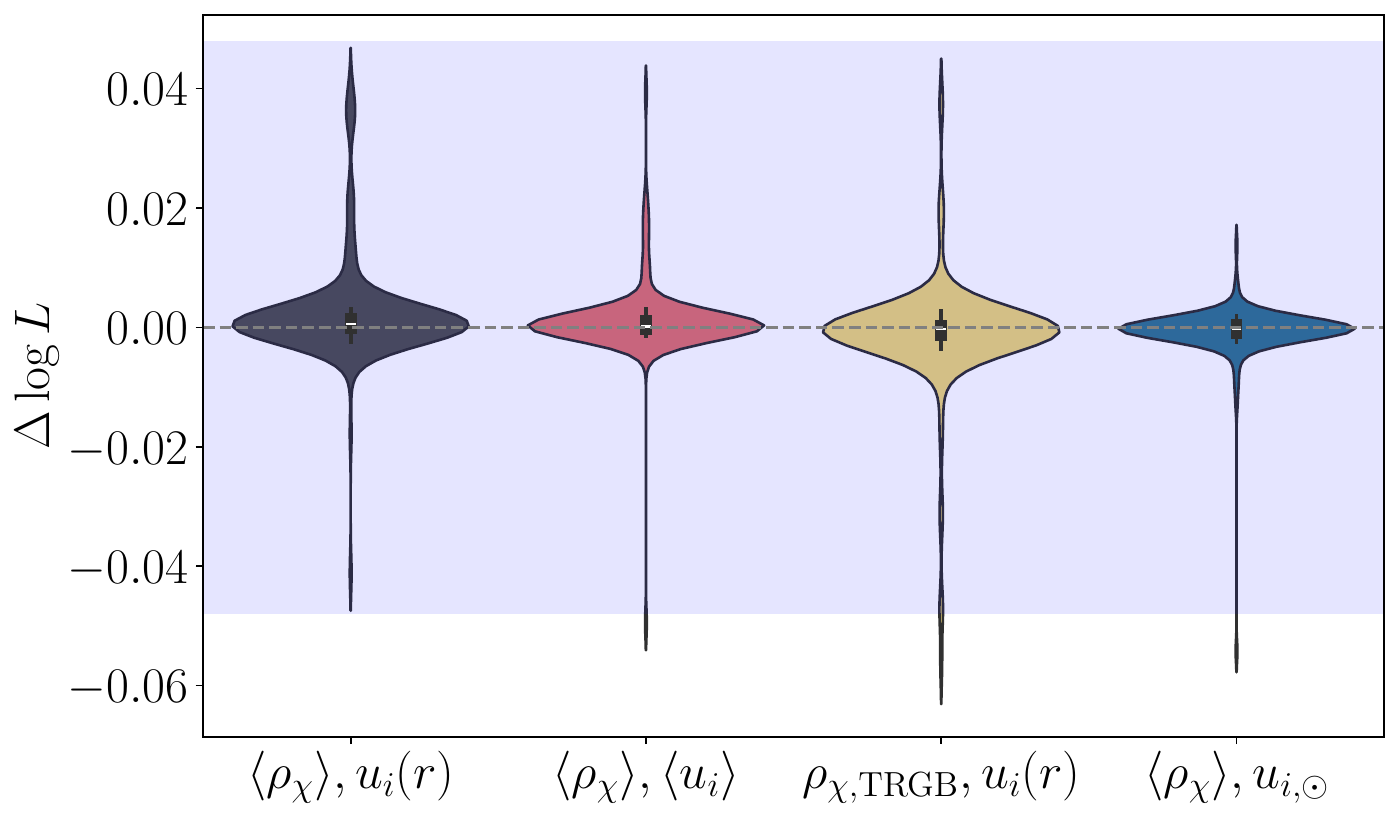}
    \caption{The effect of fixing the dark matter parameters in stellar simulations on the TRGB luminosity $\Delta \log L = \log L - \log L_{\text{full path}}$, where ``full path'' uses the full time-dependent $\rho_\chi (t)$, $u_\star(t)$ and $u_0(t)$ sampled by each globular cluster as it orbits the Galaxy.  Scenarios, from left to right, are: 1) $\rho_\chi$ is the time-averaged quantity  but $u_i = \{u_0,u_\star\}$ track the real values with time, 2) both density and the velocity parameters are time-averaged, 3) quantities are fixed to those at the presently observed location, i.e. at the TRGB value, and 4) time-averaged density, but using solar values of $u_0 = u_\star = 220$ km/s. The Lavender band represents the recommended theory uncertainty \cite{Serenelli:2017}.}
    \label{fig:violin}
\end{figure}

To quantify the effects of fixing densities and velocities, we run the following test: we simulate the full trajectory-dependent capture rate at a fixed star mass $M_\star = 0.8 M_\odot$, dark matter $m_\chi = 10$ GeV, $\sigsi = 10^{-38}$ cm$^2$ (without accounting for saturation) and metallicity $Z = 0.001$, and compare results with a few different simplifications for the parametrization of $\rho_\chi$, $u_\star$ and $u_0$ defined in Eq. \eqref{eq:fofu}. We do this for all 161 GCs with full phase-space information in the \gaia sample. We compare the change in TRGB luminosity from the full path results, and compute the difference $\Delta \log L = \log L - \log L_{\text{full path}}$. Where ``full path'' means using the full time-dependent $\rho_\chi (t)$, $u_\star(t)$ and $u_0(t)$. In Fig. \ref{fig:violin}, we present, from left to right, violin plots of  $\Delta \log L$ in the case where 1) $\rho_\chi$ is taken to be its time-averaged quantity over 5 Gyr, 2) both density and the velocity parameters are time-averaged, 3) density is fixed to the presently observed location, i.e. at the modeled star's TRGB, and 4) time-averaged density, but using solar values of $u_0 = u_\star = 220$ km/s. In all cases, the effect on the TRGB luminosity has a standard deviation well below 0.01, with very thin tails running to $\sim 0.04$. The majority of clusters are affected far less than the recommended theory error on the TRGB magnitude, shown as a lavender band, of $\Delta M_{Bol} = 0.12$ \cite{Serenelli:2017}.

We find that MESA simulations that use a value of the dark matter density equal to the average sampled value, and a dispersion and orbital speed similar to the Sun's (rightmost violin in Fig. \ref{fig:violin}), yield TRGB values that are the closest to the full trajectory. We believe that this is because averaging the density slightly overestimates capture, which is partly compensated by using larger stellar and dark matter velocities in the capture calculation. We therefore use this prescription when producing limits. Fixing the velocities also allows us to use the same results for the optical depth $\tau(r)$ \eqref{eq:opticdepthcorr} for all clusters, which significantly simplifies our analysis.

\section{Results}
\label{sec:results}
For each GC, we evolve a 0.8 $M_\odot$ star through to the TRGB, using the metallicity reported by \cite{Straniero:2020iyi} for each cluster. The chosen stellar mass value is representative of a $\sim13$ Gyr old star reaching the TRGB today, and is the same value chosen in the calibrations of Ref. \cite{Straniero:2020iyi}. We have checked that varying the stellar mass does not impact the TRGB magnitude. We refer the reader to e.g. Fig. 3 of \cite{Fung:2023euv} for a more detailed picture. We have also checked that the change due to dark matter is unchanged in a 0.9 $M_\odot$ star, for a range of dark matter masses and cross sections.

For each dark matter mass in a range from $0.1$ to $10^{4}$ GeV, we perform a set of simulations varying $\rho_\chi \sigsi$, without accounting for saturation, i.e. using Eq. \eqref{caprate} without the factor of $\eta$ \eqref{eq:extinction}.  This leads to an effective capture rate constraint. We then determine which value of $\sigsi$ leads to that capture rate when the full optical depth calculation (Eqs. \ref{eq:extinction} \& \ref{eq:opticdepthcorr}) is included. Put more simply, we map points the dashed orange line in Fig.~\ref{fig:saturation} horizontally to the blue line. If the capture rate for a given mass and cross section is above the geometric limit, then no constraint can be placed. 

For each fixed dark matter mass, we compute a test statistic: 
\begin{equation}
  TS =   -2 \left(\max_{\theta\in\Theta_0} \ell(\theta) - \max_{\theta\in\Theta} \ell(\theta)\right)
  \label{eq:TS}
\end{equation}
where $\ell$ is the log-likelihood is given by
\begin{align}
 \ell(\sigsi,\theta_1,...,\theta_n)&= \nonumber\\ 
    &\ln(N(M_\text{bol}(\sigsi,\theta_1,...,\theta_n)|M_\text{bol}^\text{obs},M_\text{bol,err}^\text{obs}))\nonumber\\ 
    &+\sum_{i=1}^n\ln(N(\theta_i|\theta^\text{obs}_{i,\text{mean}},\theta^\text{obs}_{i,\text{err}})),
\end{align} 
where $M_\text{bol}$ is the tip bolometric luminosity predicted by MESA,
 and we take each likelihood function $N(x|x^\text{obs}_\text{mean},x^\text{obs}_\text{err})$ to be a normal distribution with mean $x^\text{obs}_\text{mean}$ and standard deviation $x^\text{obs}_\text{err}$.

$\Theta$ represents the model parameter space including dark matter, and $\Theta_0$ is the standard model-only parameter space. Maximisation in Eq.~\eqref{eq:TS} is performed with Powell's method, and a root-finder is used to find values for which $TS = 2.71$, corresponding to a 90\% CL limit, assuming Wilks's theorem holds. We use the values $M_{\mathrm{bol}}^{\mathrm{obs}}$ and errors $M_{\mathrm{bol,err}}^{\mathrm{obs}}$ on the TRGB magnitude reported in \cite{Straniero:2020iyi} and shown in Fig.~\ref{fig:Mbolvsrho}, added in quadrature to the recommended theory error $\Delta M_{\mathrm{bol}} = 0.12$ from Ref. \cite{Serenelli:2017}.
As shown in Fig. 3 of Ref. \cite{Fung:2023euv}, varying the initial helium fraction $Y$ (which we set to 0.24+2Z), convective mixing length, and stellar mass within a reasonable range has very little effect on the TRGB luminosity as modeled by MESA, so we keep these fixed.

The choice dark matter distribution \textit{does} make an appreciable difference, and so both de Salas and Cautun models described in Sec. \ref{sec:potential} will be examined.
\begin{figure}
    \centering
         \hspace{-.9cm}
    \includegraphics[width=0.53\textwidth]{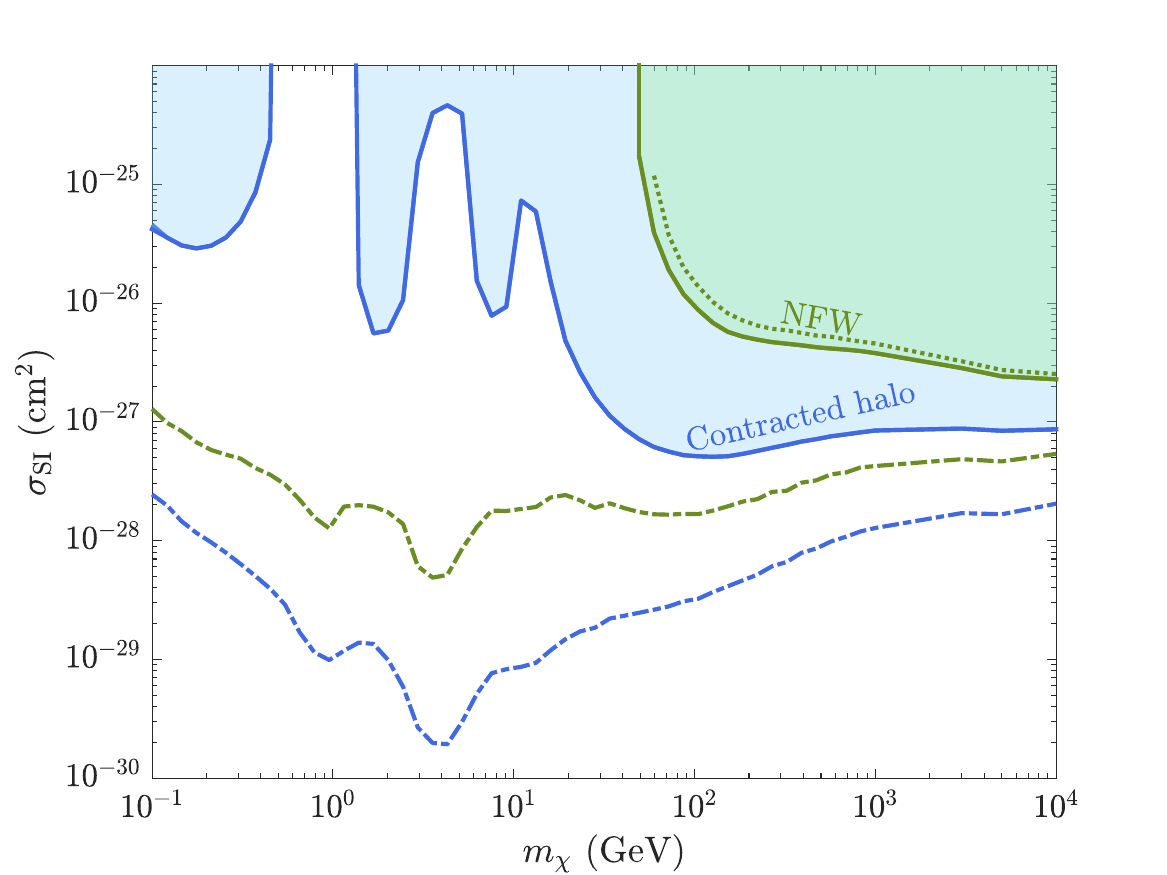}
    \caption{Limits on the dark matter-nucleon elastic scattering cross section from this work. Shaded regions are the 90\% CL bounds using the 22 globular clusters with TRGB magnitudes reported in Ref. \cite{Straniero:2020iyi}. Green: using dark matter and gravitational potential parametrized by classic NFW B1 Milky Way model from de Salas et al. \cite{2019JCAP...10..037D}. Blue: using the simulation-calibrated contracted halo model from Cautun et al. \cite{Cautun:2019eaf}. Dashed lines show projected constraints if TRGB magnitudes of all 161 clusters with \textit{Gaia} phase space data were available with errors that are 10\% of those from \cite{Straniero:2020iyi}. Dotted lines within the green region show the effects of uncertainty propagation, described in the text.}
    \label{fig:limits}
\end{figure}
Resulting 90\% CL constraints are shown in Fig.~\ref{fig:limits}. The green shaded area shows excluded dark matter parameter space derived from the de Salas NFW+B1 model. In comparison to Earth-based experiments, these limits are quite weak, coming in at best around $2 \times 10^{-27}$ cm$^2$, and losing sensitivity to dark matter masses below 40 GeV as the capture rate saturates. The stronger blue constraints use the simulation-driven baryon and contracted halo parametrization from Cautun et al. These are strongest around $m_\chi = 100$ GeV reaching $\sigsi = 5 \times 10^{-28}$ cm$^2$. Two features show a slight gain in sensitivity near 4 and 1 GeV, thanks respectively to kinematically-enhanced scattering with helium and hydrogen, coupled to a downward fluctuation in magnitude (upward fluctuation in luminosity) in the data for the clusters exposed to a larger DM flux, as seen in the rightmost points in Fig. \ref{fig:Mbolvsrho}. 

The dotted line inside the green shaded area shows the results obtained by profiling over the NFW halo shape parameters derived from $M_{200}$ and $c_{200}$, over errors on the cluster metallicities from \cite{Straniero:2020iyi}, and including the propagated errors on GC trajectories from the  \gaia  uncertainties on the 6d phase space components of each cluster. The shape parameters are drawn from Gaussian priors described by the 1d posteriors obtained on these quantities in Ref. \cite{2019JCAP...10..037D}.  We assume symmetric error bars, taking the largest error from the reported asymmetric constraints. As can be seen by comparing the dotted and solid lines, the effect of these uncertainties on our limits is minimal.

Additionally shown in Fig. \ref{fig:limits} are projected future limits assuming that  1) the TRGB luminosity could be measured for the full \textit{Gaia} sample of 161 globular clusters, with 2) measurement errors that are a factor of 10 better than errors reported in Ref. \cite{Straniero:2020iyi}. Here, we use Asimov and error bars that are 10\% of the reported TRGB magnitude error in \cite{Straniero:2020iyi}, to compute these mock constraints , which now cover the full mass range considered without saturating. The top (green) dot-dashed line shows the projection with the de Salas NFW Milky Way, and bottom (blue) corresponds to Cautun's contracted halo. The latter yields the most optimistic sensitivity, dipping near $10^{-30}$ cm$^2$ at $m_\chi = 4$ GeV. 

\section{Discussion and conclusions}
\label{sec:conclusion}
The constraints shown in Fig. \ref{fig:limits} remain far above those reported by underground direct detection experiments. Even the most optimistic sensitivity projection (dashed blue) is covered by direct detection results (most recently \cite{PandaX:2023xgl}) and sounding rocket constraints \cite{Erickcek:2007jv}, where the direct detection overburden prevents strongly-interacting particles from reaching the detector \cite{Emken:2018run}.

Nonetheless, the results present here represent a constraint that is completely independent of any systematics related to the local dark matter density and distribution, or the construction and analysis techniques used in direct detection, instead probing the properties of dark matter over the \textit{entire Galaxy}.

Prospects for improvement of these projections with statistics or measurement accuracy alone appear grim.  More sensitive probes of the stellar structure such as asteroseismology could help, though such observations in distant globular clusters are not a trivial task.

Another possibility is to measure the TRGB at locations where the dark matter density is known to be much larger than the values seen in Fig. \ref{fig:histogram}. The Galactic Centre is an obvious target, but it presents its own challenge, containing stars with a wide mix of ages and metallicities leading to a possible degeneracy with dark matter effects.

On the other hand, our results should provide some reassurance in the use of the TRGB as a standard candle in the cosmological distance ladder. A strong environmental dependence would of course confound the calibration of the next rung, and perhaps propagate to the determination of the Hubble parameter $H_0$ as in e.g. \cite{Freedman2019}. There is still some room for this to occur, but it would require a large amount of the extragalactic dataset to be from regions with dark matter densities that are orders of magnitude higher than the Milky Way's. 

\vspace{1cm}

\acknowledgements
This work was supported by the Arthur B. McDonald Canadian Astroparticle Physics Research Institute, NSERC, the Ontario Ministry of Universities and Colleges, the Canada Foundation for Innovation, and the Queen’s Centre for Advanced Computing. Research at Perimeter Institute is supported by the Government of Canada through the Department of Innovation, Science, and Economic Development, and by the Province of Ontario.  

\bibliographystyle{apsrev4-1}
\bibliography{references}

\begin{thebibliography}{73}%
\makeatletter
\providecommand \@ifxundefined [1]{%
 \@ifx{#1\undefined}
}%
\providecommand \@ifnum [1]{%
 \ifnum #1\expandafter \@firstoftwo
 \else \expandafter \@secondoftwo
 \fi
}%
\providecommand \@ifx [1]{%
 \ifx #1\expandafter \@firstoftwo
 \else \expandafter \@secondoftwo
 \fi
}%
\providecommand \natexlab [1]{#1}%
\providecommand \enquote  [1]{``#1''}%
\providecommand \bibnamefont  [1]{#1}%
\providecommand \bibfnamefont [1]{#1}%
\providecommand \citenamefont [1]{#1}%
\providecommand \href@noop [0]{\@secondoftwo}%
\providecommand \href [0]{\begingroup \@sanitize@url \@href}%
\providecommand \@href[1]{\@@startlink{#1}\@@href}%
\providecommand \@@href[1]{\endgroup#1\@@endlink}%
\providecommand \@sanitize@url [0]{\catcode `\\12\catcode `\$12\catcode
  `\&12\catcode `\#12\catcode `\^12\catcode `\_12\catcode `\%12\relax}%
\providecommand \@@startlink[1]{}%
\providecommand \@@endlink[0]{}%
\providecommand \url  [0]{\begingroup\@sanitize@url \@url }%
\providecommand \@url [1]{\endgroup\@href {#1}{\urlprefix }}%
\providecommand \urlprefix  [0]{URL }%
\providecommand \Eprint [0]{\href }%
\providecommand \doibase [0]{http://dx.doi.org/}%
\providecommand \selectlanguage [0]{\@gobble}%
\providecommand \bibinfo  [0]{\@secondoftwo}%
\providecommand \bibfield  [0]{\@secondoftwo}%
\providecommand \translation [1]{[#1]}%
\providecommand \BibitemOpen [0]{}%
\providecommand \bibitemStop [0]{}%
\providecommand \bibitemNoStop [0]{.\EOS\space}%
\providecommand \EOS [0]{\spacefactor3000\relax}%
\providecommand \BibitemShut  [1]{\csname bibitem#1\endcsname}%
\let\auto@bib@innerbib\@empty
\bibitem [{\citenamefont {Aghanim}\ \emph {et~al.}(2020)\citenamefont {Aghanim}
  \emph {et~al.}}]{Planck:2018vyg}%
  \BibitemOpen
  \bibfield  {author} {\bibinfo {author} {\bibfnamefont {N.}~\bibnamefont
  {Aghanim}} \emph {et~al.} (\bibinfo {collaboration} {Planck}),\ }\href
  {\doibase 10.1051/0004-6361/201833910} {\bibfield  {journal} {\bibinfo
  {journal} {Astron. Astrophys.}\ }\textbf {\bibinfo {volume} {641}},\ \bibinfo
  {pages} {A6} (\bibinfo {year} {2020})},\ \bibinfo {note} {[Erratum:
  Astron.Astrophys. 652, C4 (2021)]},\ \Eprint
  {http://arxiv.org/abs/1807.06209} {arXiv:1807.06209 [astro-ph.CO]}
  \BibitemShut {NoStop}%
\bibitem [{\citenamefont {Aalbers}\ \emph {et~al.}(2022)\citenamefont {Aalbers}
  \emph {et~al.}}]{LZ:2022ufs}%
  \BibitemOpen
  \bibfield  {author} {\bibinfo {author} {\bibfnamefont {J.}~\bibnamefont
  {Aalbers}} \emph {et~al.} (\bibinfo {collaboration} {LZ}),\ }\href@noop {} {\
   (\bibinfo {year} {2022})},\ \Eprint {http://arxiv.org/abs/2207.03764}
  {arXiv:2207.03764 [hep-ex]} \BibitemShut {NoStop}%
\bibitem [{\citenamefont {Aartsen}\ \emph {et~al.}(2013)\citenamefont {Aartsen}
  \emph {et~al.}}]{IceCube:2012ugg}%
  \BibitemOpen
  \bibfield  {author} {\bibinfo {author} {\bibfnamefont {M.~G.}\ \bibnamefont
  {Aartsen}} \emph {et~al.} (\bibinfo {collaboration} {IceCube}),\ }\href
  {\doibase 10.1103/PhysRevLett.110.131302} {\bibfield  {journal} {\bibinfo
  {journal} {Phys. Rev. Lett.}\ }\textbf {\bibinfo {volume} {110}},\ \bibinfo
  {pages} {131302} (\bibinfo {year} {2013})},\ \Eprint
  {http://arxiv.org/abs/1212.4097} {arXiv:1212.4097 [astro-ph.HE]} \BibitemShut
  {NoStop}%
\bibitem [{\citenamefont {Scott}\ \emph {et~al.}(2012)\citenamefont {Scott}
  \emph {et~al.}}]{IceCube:2012fvn}%
  \BibitemOpen
  \bibfield  {author} {\bibinfo {author} {\bibfnamefont {P.}~\bibnamefont
  {Scott}} \emph {et~al.} (\bibinfo {collaboration} {IceCube}),\ }\href
  {\doibase 10.1088/1475-7516/2012/11/057} {\bibfield  {journal} {\bibinfo
  {journal} {JCAP}\ }\textbf {\bibinfo {volume} {11}},\ \bibinfo {pages} {057}
  (\bibinfo {year} {2012})},\ \Eprint {http://arxiv.org/abs/1207.0810}
  {arXiv:1207.0810 [hep-ph]} \BibitemShut {NoStop}%
\bibitem [{\citenamefont {Aartsen}\ \emph {et~al.}(2016)\citenamefont {Aartsen}
  \emph {et~al.}}]{IceCube:2016yoy}%
  \BibitemOpen
  \bibfield  {author} {\bibinfo {author} {\bibfnamefont {M.~G.}\ \bibnamefont
  {Aartsen}} \emph {et~al.} (\bibinfo {collaboration} {IceCube}),\ }\href
  {\doibase 10.1088/1475-7516/2016/04/022} {\bibfield  {journal} {\bibinfo
  {journal} {JCAP}\ }\textbf {\bibinfo {volume} {04}},\ \bibinfo {pages} {022}
  (\bibinfo {year} {2016})},\ \Eprint {http://arxiv.org/abs/1601.00653}
  {arXiv:1601.00653 [hep-ph]} \BibitemShut {NoStop}%
\bibitem [{\citenamefont {Adrian-Martinez}\ \emph {et~al.}(2016)\citenamefont
  {Adrian-Martinez} \emph {et~al.}}]{ANTARES:2016xuh}%
  \BibitemOpen
  \bibfield  {author} {\bibinfo {author} {\bibfnamefont {S.}~\bibnamefont
  {Adrian-Martinez}} \emph {et~al.} (\bibinfo {collaboration} {ANTARES}),\
  }\href {\doibase 10.1016/j.physletb.2016.05.019} {\bibfield  {journal}
  {\bibinfo  {journal} {Phys. Lett. B}\ }\textbf {\bibinfo {volume} {759}},\
  \bibinfo {pages} {69} (\bibinfo {year} {2016})},\ \Eprint
  {http://arxiv.org/abs/1603.02228} {arXiv:1603.02228 [astro-ph.HE]}
  \BibitemShut {NoStop}%
\bibitem [{\citenamefont {Choi}\ \emph {et~al.}(2015)\citenamefont {Choi} \emph
  {et~al.}}]{Super-Kamiokande:2015xms}%
  \BibitemOpen
  \bibfield  {author} {\bibinfo {author} {\bibfnamefont {K.}~\bibnamefont
  {Choi}} \emph {et~al.} (\bibinfo {collaboration} {Super-Kamiokande}),\ }\href
  {\doibase 10.1103/PhysRevLett.114.141301} {\bibfield  {journal} {\bibinfo
  {journal} {Phys. Rev. Lett.}\ }\textbf {\bibinfo {volume} {114}},\ \bibinfo
  {pages} {141301} (\bibinfo {year} {2015})},\ \Eprint
  {http://arxiv.org/abs/1503.04858} {arXiv:1503.04858 [hep-ex]} \BibitemShut
  {NoStop}%
\bibitem [{\citenamefont {Raffelt}(1996)}]{Raffelt:1996wa}%
  \BibitemOpen
  \bibfield  {author} {\bibinfo {author} {\bibfnamefont {G.~G.}\ \bibnamefont
  {Raffelt}},\ }\href@noop {} {\emph {\bibinfo {title} {{Stars as laboratories
  for fundamental physics}: {The astrophysics of neutrinos, axions, and other
  weakly interacting particles}}}}\ (\bibinfo {year} {1996})\BibitemShut
  {NoStop}%
\bibitem [{\citenamefont {Press}\ and\ \citenamefont
  {Spergel}(1985)}]{Press1985CaptureParticles}%
  \BibitemOpen
  \bibfield  {author} {\bibinfo {author} {\bibfnamefont {W.~H.}\ \bibnamefont
  {Press}}\ and\ \bibinfo {author} {\bibfnamefont {D.~N.}\ \bibnamefont
  {Spergel}},\ }\href {\doibase 10.1086/163485} {\bibfield  {journal} {\bibinfo
   {journal} {\apj}\ }\textbf {\bibinfo {volume} {296}},\ \bibinfo {pages}
  {679} (\bibinfo {year} {1985})}\BibitemShut {NoStop}%
\bibitem [{\citenamefont {{Gould}}(1987{\natexlab{a}})}]{1987ApJ...321..571G}%
  \BibitemOpen
  \bibfield  {author} {\bibinfo {author} {\bibfnamefont {A.}~\bibnamefont
  {{Gould}}},\ }\href {\doibase 10.1086/165653} {\bibfield  {journal} {\bibinfo
   {journal} {\apj}\ }\textbf {\bibinfo {volume} {321}},\ \bibinfo {pages}
  {571} (\bibinfo {year} {1987}{\natexlab{a}})}\BibitemShut {NoStop}%
\bibitem [{\citenamefont {Spergel}\ and\ \citenamefont
  {Press}(1985)}]{Spergel1985EffectInterior}%
  \BibitemOpen
  \bibfield  {author} {\bibinfo {author} {\bibfnamefont {D.~N.}\ \bibnamefont
  {Spergel}}\ and\ \bibinfo {author} {\bibfnamefont {W.~H.}\ \bibnamefont
  {Press}},\ }\href {\doibase 10.1086/163336} {\bibfield  {journal} {\bibinfo
  {journal} {\apj}\ }\textbf {\bibinfo {volume} {294}},\ \bibinfo {pages} {663}
  (\bibinfo {year} {1985})}\BibitemShut {NoStop}%
\bibitem [{\citenamefont {Nauenberg}(1987)}]{Nauenberg1987}%
  \BibitemOpen
  \bibfield  {author} {\bibinfo {author} {\bibfnamefont {M.}~\bibnamefont
  {Nauenberg}},\ }\href@noop {} {\bibfield  {journal} {\bibinfo  {journal}
  {\prd}\ }\textbf {\bibinfo {volume} {36}},\ \bibinfo {pages} {1080} (\bibinfo
  {year} {1987})}\BibitemShut {NoStop}%
\bibitem [{\citenamefont {Gould}\ and\ \citenamefont
  {Raffelt}(1990)}]{Gould1990}%
  \BibitemOpen
  \bibfield  {author} {\bibinfo {author} {\bibfnamefont {A.}~\bibnamefont
  {Gould}}\ and\ \bibinfo {author} {\bibfnamefont {G.}~\bibnamefont
  {Raffelt}},\ }\href {\doibase 10.1086/168568} {\bibfield  {journal} {\bibinfo
   {journal} {\apj}\ }\textbf {\bibinfo {volume} {352}},\ \bibinfo {pages}
  {654} (\bibinfo {year} {1990})}\BibitemShut {NoStop}%
\bibitem [{\citenamefont {Lopes}\ and\ \citenamefont
  {Lopes}(2019)}]{Lopes:2019jca}%
  \BibitemOpen
  \bibfield  {author} {\bibinfo {author} {\bibfnamefont {J.}~\bibnamefont
  {Lopes}}\ and\ \bibinfo {author} {\bibfnamefont {I.}~\bibnamefont {Lopes}},\
  }\href {\doibase 10.3847/1538-4357/ab2392} {\bibfield  {journal} {\bibinfo
  {journal} {Astrophys. J.}\ }\textbf {\bibinfo {volume} {879}},\ \bibinfo
  {pages} {50} (\bibinfo {year} {2019})},\ \Eprint
  {http://arxiv.org/abs/1907.05785} {arXiv:1907.05785 [astro-ph.SR]}
  \BibitemShut {NoStop}%
\bibitem [{\citenamefont {Raen}\ \emph {et~al.}(2021)\citenamefont {Raen},
  \citenamefont {Mart\'\i{}nez-Rodr\'\i{}guez}, \citenamefont {Hurst},
  \citenamefont {Zentner}, \citenamefont {Badenes},\ and\ \citenamefont
  {Tao}}]{Raen:2020qvn}%
  \BibitemOpen
  \bibfield  {author} {\bibinfo {author} {\bibfnamefont {T.~J.}\ \bibnamefont
  {Raen}}, \bibinfo {author} {\bibfnamefont {H.}~\bibnamefont
  {Mart\'\i{}nez-Rodr\'\i{}guez}}, \bibinfo {author} {\bibfnamefont {T.~J.}\
  \bibnamefont {Hurst}}, \bibinfo {author} {\bibfnamefont {A.~R.}\ \bibnamefont
  {Zentner}}, \bibinfo {author} {\bibfnamefont {C.}~\bibnamefont {Badenes}}, \
  and\ \bibinfo {author} {\bibfnamefont {R.}~\bibnamefont {Tao}},\ }\href
  {\doibase 10.1093/mnras/stab865} {\bibfield  {journal} {\bibinfo  {journal}
  {Mon. Not. Roy. Astron. Soc.}\ }\textbf {\bibinfo {volume} {503}},\ \bibinfo
  {pages} {5611} (\bibinfo {year} {2021})},\ \Eprint
  {http://arxiv.org/abs/2010.04184} {arXiv:2010.04184 [astro-ph.GA]}
  \BibitemShut {NoStop}%
\bibitem [{\citenamefont {{Salati}}\ and\ \citenamefont
  {{Silk}}(1989)}]{SalatiSilk89}%
  \BibitemOpen
  \bibfield  {author} {\bibinfo {author} {\bibfnamefont {P.}~\bibnamefont
  {{Salati}}}\ and\ \bibinfo {author} {\bibfnamefont {J.}~\bibnamefont
  {{Silk}}},\ }\href {\doibase 10.1086/167177} {\bibfield  {journal} {\bibinfo
  {journal} {\apj}\ }\textbf {\bibinfo {volume} {338}},\ \bibinfo {pages} {24}
  (\bibinfo {year} {1989})}\BibitemShut {NoStop}%
\bibitem [{\citenamefont {{Bouquet}}\ and\ \citenamefont
  {{Salati}}(1989)}]{BouquetSalati89a}%
  \BibitemOpen
  \bibfield  {author} {\bibinfo {author} {\bibfnamefont {A.}~\bibnamefont
  {{Bouquet}}}\ and\ \bibinfo {author} {\bibfnamefont {P.}~\bibnamefont
  {{Salati}}},\ }\href@noop {} {\bibfield  {journal} {\bibinfo  {journal}
  {\aap}\ }\textbf {\bibinfo {volume} {217}},\ \bibinfo {pages} {270} (\bibinfo
  {year} {1989})}\BibitemShut {NoStop}%
\bibitem [{\citenamefont {{Moskalenko}}\ and\ \citenamefont
  {{Wai}}(2007)}]{Moskalenko07}%
  \BibitemOpen
  \bibfield  {author} {\bibinfo {author} {\bibfnamefont {I.~V.}\ \bibnamefont
  {{Moskalenko}}}\ and\ \bibinfo {author} {\bibfnamefont {L.~L.}\ \bibnamefont
  {{Wai}}},\ }\href {\doibase 10.1086/516708} {\bibfield  {journal} {\bibinfo
  {journal} {\apjl}\ }\textbf {\bibinfo {volume} {659}},\ \bibinfo {pages}
  {L29} (\bibinfo {year} {2007})},\ \Eprint
  {http://arxiv.org/abs/astro-ph/0702654} {astro-ph/0702654} \BibitemShut
  {NoStop}%
\bibitem [{\citenamefont {{Bertone}}\ and\ \citenamefont
  {{Fairbairn}}(2008)}]{Bertone07}%
  \BibitemOpen
  \bibfield  {author} {\bibinfo {author} {\bibfnamefont {G.}~\bibnamefont
  {{Bertone}}}\ and\ \bibinfo {author} {\bibfnamefont {M.}~\bibnamefont
  {{Fairbairn}}},\ }\href {\doibase 10.1103/PhysRevD.77.043515} {\bibfield
  {journal} {\bibinfo  {journal} {\prd}\ }\textbf {\bibinfo {volume} {77}},\
  \bibinfo {pages} {043515} (\bibinfo {year} {2008})},\ \Eprint
  {http://arxiv.org/abs/0709.1485} {0709.1485} \BibitemShut {NoStop}%
\bibitem [{\citenamefont {{Spolyar}}\ \emph {et~al.}(2008)\citenamefont
  {{Spolyar}}, \citenamefont {{Freese}},\ and\ \citenamefont
  {{Gondolo}}}]{Spolyar08}%
  \BibitemOpen
  \bibfield  {author} {\bibinfo {author} {\bibfnamefont {D.}~\bibnamefont
  {{Spolyar}}}, \bibinfo {author} {\bibfnamefont {K.}~\bibnamefont {{Freese}}},
  \ and\ \bibinfo {author} {\bibfnamefont {P.}~\bibnamefont {{Gondolo}}},\
  }\href {\doibase 10.1103/PhysRevLett.100.051101} {\bibfield  {journal}
  {\bibinfo  {journal} {\prl}\ }\textbf {\bibinfo {volume} {100}},\ \bibinfo
  {pages} {051101} (\bibinfo {year} {2008})},\ \Eprint
  {http://arxiv.org/abs/0705.0521} {0705.0521} \BibitemShut {NoStop}%
\bibitem [{\citenamefont {{Fairbairn}}\ \emph {et~al.}(2008)\citenamefont
  {{Fairbairn}}, \citenamefont {{Scott}},\ and\ \citenamefont
  {{Edsj{\"o}}}}]{Fairbairn08}%
  \BibitemOpen
  \bibfield  {author} {\bibinfo {author} {\bibfnamefont {M.}~\bibnamefont
  {{Fairbairn}}}, \bibinfo {author} {\bibfnamefont {P.}~\bibnamefont
  {{Scott}}}, \ and\ \bibinfo {author} {\bibfnamefont {J.}~\bibnamefont
  {{Edsj{\"o}}}},\ }\href {\doibase 10.1103/PhysRevD.77.047301} {\bibfield
  {journal} {\bibinfo  {journal} {\prd}\ }\textbf {\bibinfo {volume} {77}},\
  \bibinfo {pages} {047301} (\bibinfo {year} {2008})},\ \Eprint
  {http://arxiv.org/abs/{arXiv:0710.3396}} {{arXiv:0710.3396}} \BibitemShut
  {NoStop}%
\bibitem [{\citenamefont {{Scott}}\ \emph {et~al.}(2008)\citenamefont
  {{Scott}}, \citenamefont {{Edsj{\"o}}},\ and\ \citenamefont
  {{Fairbairn}}}]{Scott08a}%
  \BibitemOpen
  \bibfield  {author} {\bibinfo {author} {\bibfnamefont {P.}~\bibnamefont
  {{Scott}}}, \bibinfo {author} {\bibfnamefont {J.}~\bibnamefont
  {{Edsj{\"o}}}}, \ and\ \bibinfo {author} {\bibfnamefont {M.}~\bibnamefont
  {{Fairbairn}}},\ }in\ \href@noop {} {\emph {\bibinfo {booktitle} {Dark Matter
  in Astroparticle and Particle Physics: Dark 2007}}},\ \bibinfo {editor}
  {edited by\ \bibinfo {editor} {\bibfnamefont {H.~K.}\ \bibnamefont
  {{Klapdor-Kleingrothaus}}}\ and\ \bibinfo {editor} {\bibfnamefont {G.~F.}\
  \bibnamefont {{Lewis}}}}\ (\bibinfo  {publisher} {World Scientific,
  Singapore},\ \bibinfo {year} {2008})\ pp.\ \bibinfo {pages} {387--392},\
  \Eprint {http://arxiv.org/abs/{arXiv:0711.0991}} {{arXiv:0711.0991}}
  \BibitemShut {NoStop}%
\bibitem [{\citenamefont {Scott}\ \emph {et~al.}(2009)\citenamefont {Scott},
  \citenamefont {Fairbairn},\ and\ \citenamefont {Edsjo}}]{Scott:2008ns}%
  \BibitemOpen
  \bibfield  {author} {\bibinfo {author} {\bibfnamefont {P.}~\bibnamefont
  {Scott}}, \bibinfo {author} {\bibfnamefont {M.}~\bibnamefont {Fairbairn}}, \
  and\ \bibinfo {author} {\bibfnamefont {J.}~\bibnamefont {Edsjo}},\ }\href
  {\doibase 10.1111/j.1365-2966.2008.14282.x} {\bibfield  {journal} {\bibinfo
  {journal} {Mon. Not. Roy. Astron. Soc.}\ }\textbf {\bibinfo {volume} {394}},\
  \bibinfo {pages} {82} (\bibinfo {year} {2009})},\ \Eprint
  {http://arxiv.org/abs/0809.1871} {arXiv:0809.1871 [astro-ph]} \BibitemShut
  {NoStop}%
\bibitem [{\citenamefont {{Iocco}}(2008)}]{Iocco08a}%
  \BibitemOpen
  \bibfield  {author} {\bibinfo {author} {\bibfnamefont {F.}~\bibnamefont
  {{Iocco}}},\ }\href {\doibase 10.1086/587959} {\bibfield  {journal} {\bibinfo
   {journal} {\apjl}\ }\textbf {\bibinfo {volume} {677}},\ \bibinfo {pages}
  {L1} (\bibinfo {year} {2008})},\ \Eprint {http://arxiv.org/abs/0802.0941}
  {0802.0941} \BibitemShut {NoStop}%
\bibitem [{\citenamefont {{Iocco}}\ \emph {et~al.}(2008)\citenamefont
  {{Iocco}}, \citenamefont {{Bressan}}, \citenamefont {{Ripamonti}},
  \citenamefont {{Schneider}}, \citenamefont {{Ferrara}},\ and\ \citenamefont
  {{Marigo}}}]{Iocco08b}%
  \BibitemOpen
  \bibfield  {author} {\bibinfo {author} {\bibfnamefont {F.}~\bibnamefont
  {{Iocco}}}, \bibinfo {author} {\bibfnamefont {A.}~\bibnamefont {{Bressan}}},
  \bibinfo {author} {\bibfnamefont {E.}~\bibnamefont {{Ripamonti}}}, \bibinfo
  {author} {\bibfnamefont {R.}~\bibnamefont {{Schneider}}}, \bibinfo {author}
  {\bibfnamefont {A.}~\bibnamefont {{Ferrara}}}, \ and\ \bibinfo {author}
  {\bibfnamefont {P.}~\bibnamefont {{Marigo}}},\ }\href {\doibase
  10.1111/j.1365-2966.2008.13853.x} {\bibfield  {journal} {\bibinfo  {journal}
  {\mnras}\ }\textbf {\bibinfo {volume} {390}},\ \bibinfo {pages} {1655}
  (\bibinfo {year} {2008})},\ \Eprint {http://arxiv.org/abs/0805.4016}
  {arXiv:0805.4016} \BibitemShut {NoStop}%
\bibitem [{\citenamefont {{Casanellas}}\ and\ \citenamefont
  {{Lopes}}(2009)}]{Casanellas09}%
  \BibitemOpen
  \bibfield  {author} {\bibinfo {author} {\bibfnamefont {J.}~\bibnamefont
  {{Casanellas}}}\ and\ \bibinfo {author} {\bibfnamefont {I.}~\bibnamefont
  {{Lopes}}},\ }\href {\doibase 10.1088/0004-637X/705/1/135} {\bibfield
  {journal} {\bibinfo  {journal} {\apj}\ }\textbf {\bibinfo {volume} {705}},\
  \bibinfo {pages} {135} (\bibinfo {year} {2009})},\ \Eprint
  {http://arxiv.org/abs/0909.1971} {arXiv:0909.1971} \BibitemShut {NoStop}%
\bibitem [{\citenamefont {{Ripamonti}}\ \emph {et~al.}(2010)\citenamefont
  {{Ripamonti}}, \citenamefont {{Iocco}}, \citenamefont {{Ferrara}},
  \citenamefont {{Schneider}}, \citenamefont {{Bressan}},\ and\ \citenamefont
  {{Marigo}}}]{Ripamonti10}%
  \BibitemOpen
  \bibfield  {author} {\bibinfo {author} {\bibfnamefont {E.}~\bibnamefont
  {{Ripamonti}}}, \bibinfo {author} {\bibfnamefont {F.}~\bibnamefont
  {{Iocco}}}, \bibinfo {author} {\bibfnamefont {A.}~\bibnamefont {{Ferrara}}},
  \bibinfo {author} {\bibfnamefont {R.}~\bibnamefont {{Schneider}}}, \bibinfo
  {author} {\bibfnamefont {A.}~\bibnamefont {{Bressan}}}, \ and\ \bibinfo
  {author} {\bibfnamefont {P.}~\bibnamefont {{Marigo}}},\ }\href {\doibase
  10.1111/j.1365-2966.2010.16854.x} {\bibfield  {journal} {\bibinfo  {journal}
  {\mnras}\ }\textbf {\bibinfo {volume} {406}},\ \bibinfo {pages} {2605}
  (\bibinfo {year} {2010})},\ \Eprint {http://arxiv.org/abs/1003.0676}
  {arXiv:1003.0676} \BibitemShut {NoStop}%
\bibitem [{\citenamefont {{Zackrisson}}\ \emph
  {et~al.}(2010{\natexlab{a}})\citenamefont {{Zackrisson}}, \citenamefont
  {{Scott}}, \citenamefont {{Rydberg}}, \citenamefont {{Iocco}}, \citenamefont
  {{Edvardsson}}, \citenamefont {{{\"O}stlin}}, \citenamefont {{Sivertsson}},
  \citenamefont {{Zitrin}}, \citenamefont {{Broadhurst}},\ and\ \citenamefont
  {{Gondolo}}}]{Zackrisson10a}%
  \BibitemOpen
  \bibfield  {author} {\bibinfo {author} {\bibfnamefont {E.}~\bibnamefont
  {{Zackrisson}}}, \bibinfo {author} {\bibfnamefont {P.}~\bibnamefont
  {{Scott}}}, \bibinfo {author} {\bibfnamefont {C.-E.}\ \bibnamefont
  {{Rydberg}}}, \bibinfo {author} {\bibfnamefont {F.}~\bibnamefont {{Iocco}}},
  \bibinfo {author} {\bibfnamefont {B.}~\bibnamefont {{Edvardsson}}}, \bibinfo
  {author} {\bibfnamefont {G.}~\bibnamefont {{{\"O}stlin}}}, \bibinfo {author}
  {\bibfnamefont {S.}~\bibnamefont {{Sivertsson}}}, \bibinfo {author}
  {\bibfnamefont {A.}~\bibnamefont {{Zitrin}}}, \bibinfo {author}
  {\bibfnamefont {T.}~\bibnamefont {{Broadhurst}}}, \ and\ \bibinfo {author}
  {\bibfnamefont {P.}~\bibnamefont {{Gondolo}}},\ }\href {\doibase
  10.1088/0004-637X/717/1/257} {\bibfield  {journal} {\bibinfo  {journal}
  {\apj}\ }\textbf {\bibinfo {volume} {717}},\ \bibinfo {pages} {257} (\bibinfo
  {year} {2010}{\natexlab{a}})},\ \Eprint
  {http://arxiv.org/abs/{arXiv:1002.3368}} {arXiv:{arXiv:1002.3368}}
  \BibitemShut {NoStop}%
\bibitem [{\citenamefont {{Zackrisson}}\ \emph
  {et~al.}(2010{\natexlab{b}})\citenamefont {{Zackrisson}}, \citenamefont
  {{Scott}}, \citenamefont {{Rydberg}}, \citenamefont {{Iocco}}, \citenamefont
  {{Sivertsson}}, \citenamefont {{{\"O}stlin}}, \citenamefont {{Mellema}},
  \citenamefont {{Iliev}},\ and\ \citenamefont {{Shapiro}}}]{Zackrisson10b}%
  \BibitemOpen
  \bibfield  {author} {\bibinfo {author} {\bibfnamefont {E.}~\bibnamefont
  {{Zackrisson}}}, \bibinfo {author} {\bibfnamefont {P.}~\bibnamefont
  {{Scott}}}, \bibinfo {author} {\bibfnamefont {C.-E.}\ \bibnamefont
  {{Rydberg}}}, \bibinfo {author} {\bibfnamefont {F.}~\bibnamefont {{Iocco}}},
  \bibinfo {author} {\bibfnamefont {S.}~\bibnamefont {{Sivertsson}}}, \bibinfo
  {author} {\bibfnamefont {G.}~\bibnamefont {{{\"O}stlin}}}, \bibinfo {author}
  {\bibfnamefont {G.}~\bibnamefont {{Mellema}}}, \bibinfo {author}
  {\bibfnamefont {I.~T.}\ \bibnamefont {{Iliev}}}, \ and\ \bibinfo {author}
  {\bibfnamefont {P.~R.}\ \bibnamefont {{Shapiro}}},\ }\href {\doibase
  10.1111/j.1745-3933.2010.00908.x} {\bibfield  {journal} {\bibinfo  {journal}
  {\mnras}\ }\textbf {\bibinfo {volume} {407}},\ \bibinfo {pages} {L74}
  (\bibinfo {year} {2010}{\natexlab{b}})},\ \Eprint
  {http://arxiv.org/abs/{arXiv:1006.0481}} {arXiv:{arXiv:1006.0481}}
  \BibitemShut {NoStop}%
\bibitem [{\citenamefont {{Scott}}\ \emph {et~al.}(2011)\citenamefont
  {{Scott}}, \citenamefont {{Venkatesan}}, \citenamefont {{Roebber}},
  \citenamefont {{Gondolo}}, \citenamefont {{Pierpaoli}},\ and\ \citenamefont
  {{Holder}}}]{Scott11}%
  \BibitemOpen
  \bibfield  {author} {\bibinfo {author} {\bibfnamefont {P.}~\bibnamefont
  {{Scott}}}, \bibinfo {author} {\bibfnamefont {A.}~\bibnamefont
  {{Venkatesan}}}, \bibinfo {author} {\bibfnamefont {E.}~\bibnamefont
  {{Roebber}}}, \bibinfo {author} {\bibfnamefont {P.}~\bibnamefont
  {{Gondolo}}}, \bibinfo {author} {\bibfnamefont {E.}~\bibnamefont
  {{Pierpaoli}}}, \ and\ \bibinfo {author} {\bibfnamefont {G.}~\bibnamefont
  {{Holder}}},\ }\href@noop {} {\bibfield  {journal} {\bibinfo  {journal}
  {\apj}\ }\textbf {\bibinfo {volume} {742}},\ \bibinfo {pages} {129} (\bibinfo
  {year} {2011})},\ \Eprint {http://arxiv.org/abs/1107.1714} {arXiv:1107.1714}
  \BibitemShut {NoStop}%
\bibitem [{\citenamefont {Turck-Chieze}\ and\ \citenamefont
  {Lopes}(2012)}]{Turck-Chieze:2012zie}%
  \BibitemOpen
  \bibfield  {author} {\bibinfo {author} {\bibfnamefont {S.}~\bibnamefont
  {Turck-Chieze}}\ and\ \bibinfo {author} {\bibfnamefont {I.}~\bibnamefont
  {Lopes}},\ }\href {\doibase 10.1088/1674-4527/12/8/011} {\bibfield  {journal}
  {\bibinfo  {journal} {Res. Astron. Astrophys.}\ }\textbf {\bibinfo {volume}
  {12}},\ \bibinfo {pages} {1107} (\bibinfo {year} {2012})}\BibitemShut
  {NoStop}%
\bibitem [{\citenamefont {John}\ \emph {et~al.}(2024)\citenamefont {John},
  \citenamefont {Leane},\ and\ \citenamefont {Linden}}]{John:2024thz}%
  \BibitemOpen
  \bibfield  {author} {\bibinfo {author} {\bibfnamefont {I.}~\bibnamefont
  {John}}, \bibinfo {author} {\bibfnamefont {R.~K.}\ \bibnamefont {Leane}}, \
  and\ \bibinfo {author} {\bibfnamefont {T.}~\bibnamefont {Linden}},\
  }\href@noop {} {\  (\bibinfo {year} {2024})},\ \Eprint
  {http://arxiv.org/abs/2405.12267} {arXiv:2405.12267 [astro-ph.HE]}
  \BibitemShut {NoStop}%
\bibitem [{\citenamefont {Lopes}\ and\ \citenamefont
  {Lopes}(2021)}]{Lopes:2021jcy}%
  \BibitemOpen
  \bibfield  {author} {\bibinfo {author} {\bibfnamefont {J.}~\bibnamefont
  {Lopes}}\ and\ \bibinfo {author} {\bibfnamefont {I.}~\bibnamefont {Lopes}},\
  }\href {\doibase 10.1051/0004-6361/202140750} {\bibfield  {journal} {\bibinfo
   {journal} {Astron. Astrophys.}\ }\textbf {\bibinfo {volume} {651}},\
  \bibinfo {pages} {A101} (\bibinfo {year} {2021})},\ \Eprint
  {http://arxiv.org/abs/2107.13885} {arXiv:2107.13885 [astro-ph.SR]}
  \BibitemShut {NoStop}%
\bibitem [{\citenamefont {Dessert}\ and\ \citenamefont
  {Johnson}(2022)}]{Dessert:2021wjx}%
  \BibitemOpen
  \bibfield  {author} {\bibinfo {author} {\bibfnamefont {C.}~\bibnamefont
  {Dessert}}\ and\ \bibinfo {author} {\bibfnamefont {Z.}~\bibnamefont
  {Johnson}},\ }\href {\doibase 10.1103/PhysRevD.106.103034} {\bibfield
  {journal} {\bibinfo  {journal} {Phys. Rev. D}\ }\textbf {\bibinfo {volume}
  {106}},\ \bibinfo {pages} {103034} (\bibinfo {year} {2022})},\ \Eprint
  {http://arxiv.org/abs/2112.06949} {arXiv:2112.06949 [hep-ph]} \BibitemShut
  {NoStop}%
\bibitem [{\citenamefont {Kippenhahn}\ \emph {et~al.}(1990)\citenamefont
  {Kippenhahn}, \citenamefont {Weigert},\ and\ \citenamefont
  {Weiss}}]{kippenhahn1990stellar}%
  \BibitemOpen
  \bibfield  {author} {\bibinfo {author} {\bibfnamefont {R.}~\bibnamefont
  {Kippenhahn}}, \bibinfo {author} {\bibfnamefont {A.}~\bibnamefont {Weigert}},
  \ and\ \bibinfo {author} {\bibfnamefont {A.}~\bibnamefont {Weiss}},\
  }\href@noop {} {\emph {\bibinfo {title} {Stellar structure and evolution}}},\
  Vol.\ \bibinfo {volume} {192}\ (\bibinfo  {publisher} {Springer},\ \bibinfo
  {year} {1990})\BibitemShut {NoStop}%
\bibitem [{\citenamefont {{Freedman}}\ \emph {et~al.}(2019)\citenamefont
  {{Freedman}}, \citenamefont {{Madore}}, \citenamefont {{Hatt}}, \citenamefont
  {{Hoyt}}, \citenamefont {{Jang}}, \citenamefont {{Beaton}}, \citenamefont
  {{Burns}}, \citenamefont {{Lee}}, \citenamefont {{Monson}}, \citenamefont
  {{Neeley}}, \citenamefont {{Phillips}}, \citenamefont {{Rich}},\ and\
  \citenamefont {{Seibert}}}]{Freedman2019}%
  \BibitemOpen
  \bibfield  {author} {\bibinfo {author} {\bibfnamefont {W.~L.}\ \bibnamefont
  {{Freedman}}}, \bibinfo {author} {\bibfnamefont {B.~F.}\ \bibnamefont
  {{Madore}}}, \bibinfo {author} {\bibfnamefont {D.}~\bibnamefont {{Hatt}}},
  \bibinfo {author} {\bibfnamefont {T.~J.}\ \bibnamefont {{Hoyt}}}, \bibinfo
  {author} {\bibfnamefont {I.~S.}\ \bibnamefont {{Jang}}}, \bibinfo {author}
  {\bibfnamefont {R.~L.}\ \bibnamefont {{Beaton}}}, \bibinfo {author}
  {\bibfnamefont {C.~R.}\ \bibnamefont {{Burns}}}, \bibinfo {author}
  {\bibfnamefont {M.~G.}\ \bibnamefont {{Lee}}}, \bibinfo {author}
  {\bibfnamefont {A.~J.}\ \bibnamefont {{Monson}}}, \bibinfo {author}
  {\bibfnamefont {J.~R.}\ \bibnamefont {{Neeley}}}, \bibinfo {author}
  {\bibfnamefont {M.~M.}\ \bibnamefont {{Phillips}}}, \bibinfo {author}
  {\bibfnamefont {J.~A.}\ \bibnamefont {{Rich}}}, \ and\ \bibinfo {author}
  {\bibfnamefont {M.}~\bibnamefont {{Seibert}}},\ }\href {\doibase
  10.3847/1538-4357/ab2f73} {\bibfield  {journal} {\bibinfo  {journal} {\apj}\
  }\textbf {\bibinfo {volume} {882}},\ \bibinfo {eid} {34} (\bibinfo {year}
  {2019})},\ \Eprint {http://arxiv.org/abs/1907.05922} {arXiv:1907.05922
  [astro-ph.CO]} \BibitemShut {NoStop}%
\bibitem [{\citenamefont {Brown}\ \emph {et~al.}(2021)\citenamefont {Brown},
  \citenamefont {Vallenari}, \citenamefont {Prusti}, \citenamefont
  {De~Bruijne}, \citenamefont {Babusiaux}, \citenamefont {Biermann},
  \citenamefont {Creevey}, \citenamefont {Evans}, \citenamefont {Eyer},
  \citenamefont {Hutton} \emph {et~al.}}]{brown2021gaia}%
  \BibitemOpen
  \bibfield  {author} {\bibinfo {author} {\bibfnamefont {A.~G.}\ \bibnamefont
  {Brown}}, \bibinfo {author} {\bibfnamefont {A.}~\bibnamefont {Vallenari}},
  \bibinfo {author} {\bibfnamefont {T.}~\bibnamefont {Prusti}}, \bibinfo
  {author} {\bibfnamefont {J.}~\bibnamefont {De~Bruijne}}, \bibinfo {author}
  {\bibfnamefont {C.}~\bibnamefont {Babusiaux}}, \bibinfo {author}
  {\bibfnamefont {M.}~\bibnamefont {Biermann}}, \bibinfo {author}
  {\bibfnamefont {O.}~\bibnamefont {Creevey}}, \bibinfo {author} {\bibfnamefont
  {D.}~\bibnamefont {Evans}}, \bibinfo {author} {\bibfnamefont
  {L.}~\bibnamefont {Eyer}}, \bibinfo {author} {\bibfnamefont {A.}~\bibnamefont
  {Hutton}},  \emph {et~al.},\ }\href@noop {} {\bibfield  {journal} {\bibinfo
  {journal} {Astronomy \& Astrophysics}\ }\textbf {\bibinfo {volume} {649}},\
  \bibinfo {pages} {A1} (\bibinfo {year} {2021})}\BibitemShut {NoStop}%
\bibitem [{\citenamefont {{Gaia Collaboration}}(2023)}]{2023A&A...674A...1G}%
  \BibitemOpen
  \bibfield  {author} {\bibinfo {author} {\bibnamefont {{Gaia
  Collaboration}}},\ }\href {\doibase 10.1051/0004-6361/202243940} {\bibfield
  {journal} {\bibinfo  {journal} {\aap}\ }\textbf {\bibinfo {volume} {674}},\
  \bibinfo {eid} {A1} (\bibinfo {year} {2023})},\ \Eprint
  {http://arxiv.org/abs/2208.00211} {arXiv:2208.00211 [astro-ph.GA]}
  \BibitemShut {NoStop}%
\bibitem [{\citenamefont {Vasiliev}\ and\ \citenamefont
  {Baumgardt}(2021)}]{Vasiliev_2021}%
  \BibitemOpen
  \bibfield  {author} {\bibinfo {author} {\bibfnamefont {E.}~\bibnamefont
  {Vasiliev}}\ and\ \bibinfo {author} {\bibfnamefont {H.}~\bibnamefont
  {Baumgardt}},\ }\href {\doibase 10.1093/mnras/stab1475} {\bibfield  {journal}
  {\bibinfo  {journal} {Monthly Notices of the Royal Astronomical Society}\
  }\textbf {\bibinfo {volume} {505}},\ \bibinfo {pages} {5978} (\bibinfo {year}
  {2021})}\BibitemShut {NoStop}%
\bibitem [{\citenamefont {Straniero}\ \emph {et~al.}(2020)\citenamefont
  {Straniero}, \citenamefont {Pallanca}, \citenamefont {Dalessandro},
  \citenamefont {Dominguez}, \citenamefont {Ferraro}, \citenamefont
  {Giannotti}, \citenamefont {Mirizzi},\ and\ \citenamefont
  {Piersanti}}]{Straniero:2020iyi}%
  \BibitemOpen
  \bibfield  {author} {\bibinfo {author} {\bibfnamefont {O.}~\bibnamefont
  {Straniero}}, \bibinfo {author} {\bibfnamefont {C.}~\bibnamefont {Pallanca}},
  \bibinfo {author} {\bibfnamefont {E.}~\bibnamefont {Dalessandro}}, \bibinfo
  {author} {\bibfnamefont {I.}~\bibnamefont {Dominguez}}, \bibinfo {author}
  {\bibfnamefont {F.~R.}\ \bibnamefont {Ferraro}}, \bibinfo {author}
  {\bibfnamefont {M.}~\bibnamefont {Giannotti}}, \bibinfo {author}
  {\bibfnamefont {A.}~\bibnamefont {Mirizzi}}, \ and\ \bibinfo {author}
  {\bibfnamefont {L.}~\bibnamefont {Piersanti}},\ }\href {\doibase
  10.1051/0004-6361/202038775} {\bibfield  {journal} {\bibinfo  {journal}
  {Astron. Astrophys.}\ }\textbf {\bibinfo {volume} {644}},\ \bibinfo {pages}
  {A166} (\bibinfo {year} {2020})},\ \Eprint {http://arxiv.org/abs/2010.03833}
  {arXiv:2010.03833 [astro-ph.SR]} \BibitemShut {NoStop}%
\bibitem [{\citenamefont {{de Salas}}\ \emph {et~al.}(2019)\citenamefont {{de
  Salas}}, \citenamefont {{Malhan}}, \citenamefont {{Freese}}, \citenamefont
  {{Hattori}},\ and\ \citenamefont {{Valluri}}}]{2019JCAP...10..037D}%
  \BibitemOpen
  \bibfield  {author} {\bibinfo {author} {\bibfnamefont {P.~F.}\ \bibnamefont
  {{de Salas}}}, \bibinfo {author} {\bibfnamefont {K.}~\bibnamefont
  {{Malhan}}}, \bibinfo {author} {\bibfnamefont {K.}~\bibnamefont {{Freese}}},
  \bibinfo {author} {\bibfnamefont {K.}~\bibnamefont {{Hattori}}}, \ and\
  \bibinfo {author} {\bibfnamefont {M.}~\bibnamefont {{Valluri}}},\ }\href
  {\doibase 10.1088/1475-7516/2019/10/037} {\bibfield  {journal} {\bibinfo
  {journal} {\jcap}\ }\textbf {\bibinfo {volume} {2019}},\ \bibinfo {eid} {037}
  (\bibinfo {year} {2019})},\ \Eprint {http://arxiv.org/abs/1906.06133}
  {arXiv:1906.06133 [astro-ph.GA]} \BibitemShut {NoStop}%
\bibitem [{\citenamefont {Cautun}\ \emph {et~al.}(2020)\citenamefont {Cautun},
  \citenamefont {Benitez-Llambay}, \citenamefont {Deason}, \citenamefont
  {Frenk}, \citenamefont {Fattahi}, \citenamefont {G\'omez}, \citenamefont
  {Grand}, \citenamefont {Oman}, \citenamefont {Navarro},\ and\ \citenamefont
  {Simpson}}]{Cautun:2019eaf}%
  \BibitemOpen
  \bibfield  {author} {\bibinfo {author} {\bibfnamefont {M.}~\bibnamefont
  {Cautun}}, \bibinfo {author} {\bibfnamefont {A.}~\bibnamefont
  {Benitez-Llambay}}, \bibinfo {author} {\bibfnamefont {A.~J.}\ \bibnamefont
  {Deason}}, \bibinfo {author} {\bibfnamefont {C.~S.}\ \bibnamefont {Frenk}},
  \bibinfo {author} {\bibfnamefont {A.}~\bibnamefont {Fattahi}}, \bibinfo
  {author} {\bibfnamefont {F.~A.}\ \bibnamefont {G\'omez}}, \bibinfo {author}
  {\bibfnamefont {R.~J.~J.}\ \bibnamefont {Grand}}, \bibinfo {author}
  {\bibfnamefont {K.~A.}\ \bibnamefont {Oman}}, \bibinfo {author}
  {\bibfnamefont {J.~F.}\ \bibnamefont {Navarro}}, \ and\ \bibinfo {author}
  {\bibfnamefont {C.~M.}\ \bibnamefont {Simpson}},\ }\href {\doibase
  10.1093/mnras/staa1017} {\bibfield  {journal} {\bibinfo  {journal} {Mon. Not.
  Roy. Astron. Soc.}\ }\textbf {\bibinfo {volume} {494}},\ \bibinfo {pages}
  {4291} (\bibinfo {year} {2020})},\ \Eprint {http://arxiv.org/abs/1911.04557}
  {arXiv:1911.04557 [astro-ph.GA]} \BibitemShut {NoStop}%
\bibitem [{\citenamefont {{Eilers}}\ \emph {et~al.}(2019)\citenamefont
  {{Eilers}}, \citenamefont {{Hogg}}, \citenamefont {{Rix}},\ and\
  \citenamefont {{Ness}}}]{2019ApJ...871..120E}%
  \BibitemOpen
  \bibfield  {author} {\bibinfo {author} {\bibfnamefont {A.-C.}\ \bibnamefont
  {{Eilers}}}, \bibinfo {author} {\bibfnamefont {D.~W.}\ \bibnamefont
  {{Hogg}}}, \bibinfo {author} {\bibfnamefont {H.-W.}\ \bibnamefont {{Rix}}}, \
  and\ \bibinfo {author} {\bibfnamefont {M.~K.}\ \bibnamefont {{Ness}}},\
  }\href {\doibase 10.3847/1538-4357/aaf648} {\bibfield  {journal} {\bibinfo
  {journal} {\apj}\ }\textbf {\bibinfo {volume} {871}},\ \bibinfo {eid} {120}
  (\bibinfo {year} {2019})},\ \Eprint {http://arxiv.org/abs/1810.09466}
  {arXiv:1810.09466 [astro-ph.GA]} \BibitemShut {NoStop}%
\bibitem [{\citenamefont {{Miyamoto}}\ and\ \citenamefont
  {{Nagai}}(1975)}]{1975PASJ...27..533M}%
  \BibitemOpen
  \bibfield  {author} {\bibinfo {author} {\bibfnamefont {M.}~\bibnamefont
  {{Miyamoto}}}\ and\ \bibinfo {author} {\bibfnamefont {R.}~\bibnamefont
  {{Nagai}}},\ }\href@noop {} {\bibfield  {journal} {\bibinfo  {journal}
  {\pasj}\ }\textbf {\bibinfo {volume} {27}},\ \bibinfo {pages} {533} (\bibinfo
  {year} {1975})}\BibitemShut {NoStop}%
\bibitem [{\citenamefont {{Grand}}\ \emph {et~al.}(2017)\citenamefont
  {{Grand}}, \citenamefont {{G{\'o}mez}}, \citenamefont {{Marinacci}},
  \citenamefont {{Pakmor}}, \citenamefont {{Springel}}, \citenamefont
  {{Campbell}}, \citenamefont {{Frenk}}, \citenamefont {{Jenkins}},\ and\
  \citenamefont {{White}}}]{2017MNRAS.467..179G}%
  \BibitemOpen
  \bibfield  {author} {\bibinfo {author} {\bibfnamefont {R.~J.~J.}\
  \bibnamefont {{Grand}}}, \bibinfo {author} {\bibfnamefont {F.~A.}\
  \bibnamefont {{G{\'o}mez}}}, \bibinfo {author} {\bibfnamefont
  {F.}~\bibnamefont {{Marinacci}}}, \bibinfo {author} {\bibfnamefont
  {R.}~\bibnamefont {{Pakmor}}}, \bibinfo {author} {\bibfnamefont
  {V.}~\bibnamefont {{Springel}}}, \bibinfo {author} {\bibfnamefont {D.~J.~R.}\
  \bibnamefont {{Campbell}}}, \bibinfo {author} {\bibfnamefont {C.~S.}\
  \bibnamefont {{Frenk}}}, \bibinfo {author} {\bibfnamefont {A.}~\bibnamefont
  {{Jenkins}}}, \ and\ \bibinfo {author} {\bibfnamefont {S.~D.~M.}\
  \bibnamefont {{White}}},\ }\href {\doibase 10.1093/mnras/stx071} {\bibfield
  {journal} {\bibinfo  {journal} {\mnras}\ }\textbf {\bibinfo {volume} {467}},\
  \bibinfo {pages} {179} (\bibinfo {year} {2017})},\ \Eprint
  {http://arxiv.org/abs/1610.01159} {arXiv:1610.01159 [astro-ph.GA]}
  \BibitemShut {NoStop}%
\bibitem [{\citenamefont {{Fattahi}}\ \emph {et~al.}(2016)\citenamefont
  {{Fattahi}}, \citenamefont {{Navarro}}, \citenamefont {{Sawala}},
  \citenamefont {{Frenk}}, \citenamefont {{Oman}}, \citenamefont {{Crain}},
  \citenamefont {{Furlong}}, \citenamefont {{Schaller}}, \citenamefont
  {{Schaye}}, \citenamefont {{Theuns}},\ and\ \citenamefont
  {{Jenkins}}}]{2016MNRAS.457..844F}%
  \BibitemOpen
  \bibfield  {author} {\bibinfo {author} {\bibfnamefont {A.}~\bibnamefont
  {{Fattahi}}}, \bibinfo {author} {\bibfnamefont {J.~F.}\ \bibnamefont
  {{Navarro}}}, \bibinfo {author} {\bibfnamefont {T.}~\bibnamefont {{Sawala}}},
  \bibinfo {author} {\bibfnamefont {C.~S.}\ \bibnamefont {{Frenk}}}, \bibinfo
  {author} {\bibfnamefont {K.~A.}\ \bibnamefont {{Oman}}}, \bibinfo {author}
  {\bibfnamefont {R.~A.}\ \bibnamefont {{Crain}}}, \bibinfo {author}
  {\bibfnamefont {M.}~\bibnamefont {{Furlong}}}, \bibinfo {author}
  {\bibfnamefont {M.}~\bibnamefont {{Schaller}}}, \bibinfo {author}
  {\bibfnamefont {J.}~\bibnamefont {{Schaye}}}, \bibinfo {author}
  {\bibfnamefont {T.}~\bibnamefont {{Theuns}}}, \ and\ \bibinfo {author}
  {\bibfnamefont {A.}~\bibnamefont {{Jenkins}}},\ }\href {\doibase
  10.1093/mnras/stv2970} {\bibfield  {journal} {\bibinfo  {journal} {\mnras}\
  }\textbf {\bibinfo {volume} {457}},\ \bibinfo {pages} {844} (\bibinfo {year}
  {2016})},\ \Eprint {http://arxiv.org/abs/1507.03643} {arXiv:1507.03643
  [astro-ph.GA]} \BibitemShut {NoStop}%
\bibitem [{\citenamefont {{Sawala}}\ \emph {et~al.}(2016)\citenamefont
  {{Sawala}}, \citenamefont {{Frenk}}, \citenamefont {{Fattahi}}, \citenamefont
  {{Navarro}}, \citenamefont {{Bower}}, \citenamefont {{Crain}}, \citenamefont
  {{Dalla Vecchia}}, \citenamefont {{Furlong}}, \citenamefont {{Helly}},
  \citenamefont {{Jenkins}}, \citenamefont {{Oman}}, \citenamefont
  {{Schaller}}, \citenamefont {{Schaye}}, \citenamefont {{Theuns}},
  \citenamefont {{Trayford}},\ and\ \citenamefont
  {{White}}}]{2016MNRAS.457.1931S}%
  \BibitemOpen
  \bibfield  {author} {\bibinfo {author} {\bibfnamefont {T.}~\bibnamefont
  {{Sawala}}}, \bibinfo {author} {\bibfnamefont {C.~S.}\ \bibnamefont
  {{Frenk}}}, \bibinfo {author} {\bibfnamefont {A.}~\bibnamefont {{Fattahi}}},
  \bibinfo {author} {\bibfnamefont {J.~F.}\ \bibnamefont {{Navarro}}}, \bibinfo
  {author} {\bibfnamefont {R.~G.}\ \bibnamefont {{Bower}}}, \bibinfo {author}
  {\bibfnamefont {R.~A.}\ \bibnamefont {{Crain}}}, \bibinfo {author}
  {\bibfnamefont {C.}~\bibnamefont {{Dalla Vecchia}}}, \bibinfo {author}
  {\bibfnamefont {M.}~\bibnamefont {{Furlong}}}, \bibinfo {author}
  {\bibfnamefont {J.~C.}\ \bibnamefont {{Helly}}}, \bibinfo {author}
  {\bibfnamefont {A.}~\bibnamefont {{Jenkins}}}, \bibinfo {author}
  {\bibfnamefont {K.~A.}\ \bibnamefont {{Oman}}}, \bibinfo {author}
  {\bibfnamefont {M.}~\bibnamefont {{Schaller}}}, \bibinfo {author}
  {\bibfnamefont {J.}~\bibnamefont {{Schaye}}}, \bibinfo {author}
  {\bibfnamefont {T.}~\bibnamefont {{Theuns}}}, \bibinfo {author}
  {\bibfnamefont {J.}~\bibnamefont {{Trayford}}}, \ and\ \bibinfo {author}
  {\bibfnamefont {S.~D.~M.}\ \bibnamefont {{White}}},\ }\href {\doibase
  10.1093/mnras/stw145} {\bibfield  {journal} {\bibinfo  {journal} {\mnras}\
  }\textbf {\bibinfo {volume} {457}},\ \bibinfo {pages} {1931} (\bibinfo {year}
  {2016})},\ \Eprint {http://arxiv.org/abs/1511.01098} {arXiv:1511.01098
  [astro-ph.GA]} \BibitemShut {NoStop}%
\bibitem [{\citenamefont {{Schaye}}\ \emph {et~al.}(2015)\citenamefont
  {{Schaye}}, \citenamefont {{Crain}}, \citenamefont {{Bower}}, \citenamefont
  {{Furlong}}, \citenamefont {{Schaller}}, \citenamefont {{Theuns}},
  \citenamefont {{Dalla Vecchia}}, \citenamefont {{Frenk}}, \citenamefont
  {{McCarthy}}, \citenamefont {{Helly}}, \citenamefont {{Jenkins}},
  \citenamefont {{Rosas-Guevara}}, \citenamefont {{White}}, \citenamefont
  {{Baes}}, \citenamefont {{Booth}}, \citenamefont {{Camps}}, \citenamefont
  {{Navarro}}, \citenamefont {{Qu}}, \citenamefont {{Rahmati}}, \citenamefont
  {{Sawala}}, \citenamefont {{Thomas}},\ and\ \citenamefont
  {{Trayford}}}]{2015MNRAS.446..521S}%
  \BibitemOpen
  \bibfield  {author} {\bibinfo {author} {\bibfnamefont {J.}~\bibnamefont
  {{Schaye}}}, \bibinfo {author} {\bibfnamefont {R.~A.}\ \bibnamefont
  {{Crain}}}, \bibinfo {author} {\bibfnamefont {R.~G.}\ \bibnamefont
  {{Bower}}}, \bibinfo {author} {\bibfnamefont {M.}~\bibnamefont {{Furlong}}},
  \bibinfo {author} {\bibfnamefont {M.}~\bibnamefont {{Schaller}}}, \bibinfo
  {author} {\bibfnamefont {T.}~\bibnamefont {{Theuns}}}, \bibinfo {author}
  {\bibfnamefont {C.}~\bibnamefont {{Dalla Vecchia}}}, \bibinfo {author}
  {\bibfnamefont {C.~S.}\ \bibnamefont {{Frenk}}}, \bibinfo {author}
  {\bibfnamefont {I.~G.}\ \bibnamefont {{McCarthy}}}, \bibinfo {author}
  {\bibfnamefont {J.~C.}\ \bibnamefont {{Helly}}}, \bibinfo {author}
  {\bibfnamefont {A.}~\bibnamefont {{Jenkins}}}, \bibinfo {author}
  {\bibfnamefont {Y.~M.}\ \bibnamefont {{Rosas-Guevara}}}, \bibinfo {author}
  {\bibfnamefont {S.~D.~M.}\ \bibnamefont {{White}}}, \bibinfo {author}
  {\bibfnamefont {M.}~\bibnamefont {{Baes}}}, \bibinfo {author} {\bibfnamefont
  {C.~M.}\ \bibnamefont {{Booth}}}, \bibinfo {author} {\bibfnamefont
  {P.}~\bibnamefont {{Camps}}}, \bibinfo {author} {\bibfnamefont {J.~F.}\
  \bibnamefont {{Navarro}}}, \bibinfo {author} {\bibfnamefont {Y.}~\bibnamefont
  {{Qu}}}, \bibinfo {author} {\bibfnamefont {A.}~\bibnamefont {{Rahmati}}},
  \bibinfo {author} {\bibfnamefont {T.}~\bibnamefont {{Sawala}}}, \bibinfo
  {author} {\bibfnamefont {P.~A.}\ \bibnamefont {{Thomas}}}, \ and\ \bibinfo
  {author} {\bibfnamefont {J.}~\bibnamefont {{Trayford}}},\ }\href {\doibase
  10.1093/mnras/stu2058} {\bibfield  {journal} {\bibinfo  {journal} {\mnras}\
  }\textbf {\bibinfo {volume} {446}},\ \bibinfo {pages} {521} (\bibinfo {year}
  {2015})},\ \Eprint {http://arxiv.org/abs/1407.7040} {arXiv:1407.7040
  [astro-ph.GA]} \BibitemShut {NoStop}%
\bibitem [{\citenamefont {{McMillan}}(2017)}]{2017MNRAS.465...76M}%
  \BibitemOpen
  \bibfield  {author} {\bibinfo {author} {\bibfnamefont {P.~J.}\ \bibnamefont
  {{McMillan}}},\ }\href {\doibase 10.1093/mnras/stw2759} {\bibfield  {journal}
  {\bibinfo  {journal} {\mnras}\ }\textbf {\bibinfo {volume} {465}},\ \bibinfo
  {pages} {76} (\bibinfo {year} {2017})},\ \Eprint
  {http://arxiv.org/abs/1608.00971} {arXiv:1608.00971 [astro-ph.GA]}
  \BibitemShut {NoStop}%
\bibitem [{\citenamefont {{Bovy}}(2015)}]{2015ApJS..216...29B}%
  \BibitemOpen
  \bibfield  {author} {\bibinfo {author} {\bibfnamefont {J.}~\bibnamefont
  {{Bovy}}},\ }\href {\doibase 10.1088/0067-0049/216/2/29} {\bibfield
  {journal} {\bibinfo  {journal} {\apjs}\ }\textbf {\bibinfo {volume} {216}},\
  \bibinfo {eid} {29} (\bibinfo {year} {2015})},\ \Eprint
  {http://arxiv.org/abs/1412.3451} {arXiv:1412.3451 [astro-ph.GA]} \BibitemShut
  {NoStop}%
\bibitem [{\citenamefont {Binney}\ and\ \citenamefont
  {Tremaine}(2011)}]{binney2011galactic}%
  \BibitemOpen
  \bibfield  {author} {\bibinfo {author} {\bibfnamefont {J.}~\bibnamefont
  {Binney}}\ and\ \bibinfo {author} {\bibfnamefont {S.}~\bibnamefont
  {Tremaine}},\ }\href@noop {} {\emph {\bibinfo {title} {Galactic dynamics}}},\
  Vol.~\bibinfo {volume} {13}\ (\bibinfo  {publisher} {Princeton university
  press},\ \bibinfo {year} {2011})\BibitemShut {NoStop}%
\bibitem [{\citenamefont {Dutton}\ and\ \citenamefont
  {Macci\`o}(2014)}]{Dutton:2014xda}%
  \BibitemOpen
  \bibfield  {author} {\bibinfo {author} {\bibfnamefont {A.~A.}\ \bibnamefont
  {Dutton}}\ and\ \bibinfo {author} {\bibfnamefont {A.~V.}\ \bibnamefont
  {Macci\`o}},\ }\href {\doibase 10.1093/mnras/stu742} {\bibfield  {journal}
  {\bibinfo  {journal} {Mon. Not. Roy. Astron. Soc.}\ }\textbf {\bibinfo
  {volume} {441}},\ \bibinfo {pages} {3359} (\bibinfo {year} {2014})},\ \Eprint
  {http://arxiv.org/abs/1402.7073} {arXiv:1402.7073 [astro-ph.CO]} \BibitemShut
  {NoStop}%
\bibitem [{\citenamefont {{Gould}}(1987{\natexlab{b}})}]{Gould87a}%
  \BibitemOpen
  \bibfield  {author} {\bibinfo {author} {\bibfnamefont {A.}~\bibnamefont
  {{Gould}}},\ }\href {\doibase 10.1086/165652} {\bibfield  {journal} {\bibinfo
   {journal} {\apj}\ }\textbf {\bibinfo {volume} {321}},\ \bibinfo {pages}
  {560} (\bibinfo {year} {1987}{\natexlab{b}})}\BibitemShut {NoStop}%
\bibitem [{\citenamefont {Moore}(1996)}]{Moore:1995pb}%
  \BibitemOpen
  \bibfield  {author} {\bibinfo {author} {\bibfnamefont {B.}~\bibnamefont
  {Moore}},\ }\href {\doibase 10.1086/309998} {\bibfield  {journal} {\bibinfo
  {journal} {Astrophys. J. Lett.}\ }\textbf {\bibinfo {volume} {461}},\
  \bibinfo {pages} {L13} (\bibinfo {year} {1996})},\ \Eprint
  {http://arxiv.org/abs/astro-ph/9511147} {arXiv:astro-ph/9511147} \BibitemShut
  {NoStop}%
\bibitem [{\citenamefont {Saitoh}\ \emph {et~al.}(2006)\citenamefont {Saitoh},
  \citenamefont {Koda}, \citenamefont {Okamoto}, \citenamefont {Wada},\ and\
  \citenamefont {Habe}}]{Saitoh:2005tt}%
  \BibitemOpen
  \bibfield  {author} {\bibinfo {author} {\bibfnamefont {T.~R.}\ \bibnamefont
  {Saitoh}}, \bibinfo {author} {\bibfnamefont {J.}~\bibnamefont {Koda}},
  \bibinfo {author} {\bibfnamefont {T.}~\bibnamefont {Okamoto}}, \bibinfo
  {author} {\bibfnamefont {K.}~\bibnamefont {Wada}}, \ and\ \bibinfo {author}
  {\bibfnamefont {A.}~\bibnamefont {Habe}},\ }\href {\doibase 10.1086/500104}
  {\bibfield  {journal} {\bibinfo  {journal} {Astrophys. J.}\ }\textbf
  {\bibinfo {volume} {640}},\ \bibinfo {pages} {22} (\bibinfo {year} {2006})},\
  \Eprint {http://arxiv.org/abs/astro-ph/0511692} {arXiv:astro-ph/0511692}
  \BibitemShut {NoStop}%
\bibitem [{\citenamefont {Bramante}\ \emph {et~al.}(2017)\citenamefont
  {Bramante}, \citenamefont {Delgado},\ and\ \citenamefont
  {Martin}}]{Bramante:2017xlb}%
  \BibitemOpen
  \bibfield  {author} {\bibinfo {author} {\bibfnamefont {J.}~\bibnamefont
  {Bramante}}, \bibinfo {author} {\bibfnamefont {A.}~\bibnamefont {Delgado}}, \
  and\ \bibinfo {author} {\bibfnamefont {A.}~\bibnamefont {Martin}},\ }\href
  {\doibase 10.1103/PhysRevD.96.063002} {\bibfield  {journal} {\bibinfo
  {journal} {Phys. Rev. D}\ }\textbf {\bibinfo {volume} {96}},\ \bibinfo
  {pages} {063002} (\bibinfo {year} {2017})},\ \Eprint
  {http://arxiv.org/abs/1703.04043} {arXiv:1703.04043 [hep-ph]} \BibitemShut
  {NoStop}%
\bibitem [{\citenamefont {Leane}\ and\ \citenamefont
  {Smirnov}(2023)}]{Leane:2023woh}%
  \BibitemOpen
  \bibfield  {author} {\bibinfo {author} {\bibfnamefont {R.~K.}\ \bibnamefont
  {Leane}}\ and\ \bibinfo {author} {\bibfnamefont {J.}~\bibnamefont
  {Smirnov}},\ }\href {\doibase 10.1088/1475-7516/2023/12/040} {\bibfield
  {journal} {\bibinfo  {journal} {JCAP}\ }\textbf {\bibinfo {volume} {12}},\
  \bibinfo {pages} {040} (\bibinfo {year} {2023})},\ \Eprint
  {http://arxiv.org/abs/2309.00669} {arXiv:2309.00669 [hep-ph]} \BibitemShut
  {NoStop}%
\bibitem [{\citenamefont {Ilie}(2024)}]{Ilie:2024sos}%
  \BibitemOpen
  \bibfield  {author} {\bibinfo {author} {\bibfnamefont {C.}~\bibnamefont
  {Ilie}},\ }\href@noop {} {\  (\bibinfo {year} {2024})},\ \Eprint
  {http://arxiv.org/abs/2402.07713} {arXiv:2402.07713 [astro-ph.CO]}
  \BibitemShut {NoStop}%
\bibitem [{\citenamefont {Bell}\ \emph {et~al.}(2024)\citenamefont {Bell},
  \citenamefont {Busoni}, \citenamefont {Robles},\ and\ \citenamefont
  {Virgato}}]{Bell:2024qmj}%
  \BibitemOpen
  \bibfield  {author} {\bibinfo {author} {\bibfnamefont {N.~F.}\ \bibnamefont
  {Bell}}, \bibinfo {author} {\bibfnamefont {G.}~\bibnamefont {Busoni}},
  \bibinfo {author} {\bibfnamefont {S.}~\bibnamefont {Robles}}, \ and\ \bibinfo
  {author} {\bibfnamefont {M.}~\bibnamefont {Virgato}},\ }\href@noop {} {\
  (\bibinfo {year} {2024})},\ \Eprint {http://arxiv.org/abs/2404.16272}
  {arXiv:2404.16272 [hep-ph]} \BibitemShut {NoStop}%
\bibitem [{\citenamefont {{Gould}}(1992)}]{Gould92OpticalDepth}%
  \BibitemOpen
  \bibfield  {author} {\bibinfo {author} {\bibfnamefont {A.}~\bibnamefont
  {{Gould}}},\ }\href {\doibase 10.1086/171057} {\bibfield  {journal} {\bibinfo
   {journal} {\apj}\ }\textbf {\bibinfo {volume} {387}},\ \bibinfo {pages} {21}
  (\bibinfo {year} {1992})}\BibitemShut {NoStop}%
\bibitem [{\citenamefont {Busoni}\ \emph {et~al.}(2017)\citenamefont {Busoni},
  \citenamefont {De~Simone}, \citenamefont {Scott},\ and\ \citenamefont
  {Vincent}}]{Busoni:2017mhe}%
  \BibitemOpen
  \bibfield  {author} {\bibinfo {author} {\bibfnamefont {G.}~\bibnamefont
  {Busoni}}, \bibinfo {author} {\bibfnamefont {A.}~\bibnamefont {De~Simone}},
  \bibinfo {author} {\bibfnamefont {P.}~\bibnamefont {Scott}}, \ and\ \bibinfo
  {author} {\bibfnamefont {A.~C.}\ \bibnamefont {Vincent}},\ }\href {\doibase
  10.1088/1475-7516/2017/10/037} {\bibfield  {journal} {\bibinfo  {journal}
  {JCAP}\ }\textbf {\bibinfo {volume} {10}},\ \bibinfo {pages} {037} (\bibinfo
  {year} {2017})},\ \Eprint {http://arxiv.org/abs/1703.07784} {arXiv:1703.07784
  [hep-ph]} \BibitemShut {NoStop}%
\bibitem [{\citenamefont {Garani}\ and\ \citenamefont
  {Palomares-Ruiz}(2022)}]{Garani:2021feo}%
  \BibitemOpen
  \bibfield  {author} {\bibinfo {author} {\bibfnamefont {R.}~\bibnamefont
  {Garani}}\ and\ \bibinfo {author} {\bibfnamefont {S.}~\bibnamefont
  {Palomares-Ruiz}},\ }\href {\doibase 10.1088/1475-7516/2022/05/042}
  {\bibfield  {journal} {\bibinfo  {journal} {JCAP}\ }\textbf {\bibinfo
  {volume} {05}},\ \bibinfo {pages} {042} (\bibinfo {year} {2022})},\ \Eprint
  {http://arxiv.org/abs/2104.12757} {arXiv:2104.12757 [hep-ph]} \BibitemShut
  {NoStop}%
\bibitem [{\citenamefont {{Paxton}}\ \emph {et~al.}(2011)\citenamefont
  {{Paxton}}, \citenamefont {{Bildsten}}, \citenamefont {{Dotter}},
  \citenamefont {{Herwig}}, \citenamefont {{Lesaffre}},\ and\ \citenamefont
  {{Timmes}}}]{2011ApJS..192....3P}%
  \BibitemOpen
  \bibfield  {author} {\bibinfo {author} {\bibfnamefont {B.}~\bibnamefont
  {{Paxton}}}, \bibinfo {author} {\bibfnamefont {L.}~\bibnamefont
  {{Bildsten}}}, \bibinfo {author} {\bibfnamefont {A.}~\bibnamefont
  {{Dotter}}}, \bibinfo {author} {\bibfnamefont {F.}~\bibnamefont {{Herwig}}},
  \bibinfo {author} {\bibfnamefont {P.}~\bibnamefont {{Lesaffre}}}, \ and\
  \bibinfo {author} {\bibfnamefont {F.}~\bibnamefont {{Timmes}}},\ }\href
  {\doibase 10.1088/0067-0049/192/1/3} {\bibfield  {journal} {\bibinfo
  {journal} {\apjs}\ }\textbf {\bibinfo {volume} {192}},\ \bibinfo {eid} {3}
  (\bibinfo {year} {2011})},\ \Eprint {http://arxiv.org/abs/1009.1622}
  {arXiv:1009.1622 [astro-ph.SR]} \BibitemShut {NoStop}%
\bibitem [{\citenamefont {{Paxton}}\ \emph {et~al.}(2013)\citenamefont
  {{Paxton}}, \citenamefont {{Cantiello}}, \citenamefont {{Arras}},
  \citenamefont {{Bildsten}}, \citenamefont {{Brown}}, \citenamefont
  {{Dotter}}, \citenamefont {{Mankovich}}, \citenamefont {{Montgomery}},
  \citenamefont {{Stello}}, \citenamefont {{Timmes}},\ and\ \citenamefont
  {{Townsend}}}]{2013ApJS..208....4P}%
  \BibitemOpen
  \bibfield  {author} {\bibinfo {author} {\bibfnamefont {B.}~\bibnamefont
  {{Paxton}}}, \bibinfo {author} {\bibfnamefont {M.}~\bibnamefont
  {{Cantiello}}}, \bibinfo {author} {\bibfnamefont {P.}~\bibnamefont
  {{Arras}}}, \bibinfo {author} {\bibfnamefont {L.}~\bibnamefont {{Bildsten}}},
  \bibinfo {author} {\bibfnamefont {E.~F.}\ \bibnamefont {{Brown}}}, \bibinfo
  {author} {\bibfnamefont {A.}~\bibnamefont {{Dotter}}}, \bibinfo {author}
  {\bibfnamefont {C.}~\bibnamefont {{Mankovich}}}, \bibinfo {author}
  {\bibfnamefont {M.~H.}\ \bibnamefont {{Montgomery}}}, \bibinfo {author}
  {\bibfnamefont {D.}~\bibnamefont {{Stello}}}, \bibinfo {author}
  {\bibfnamefont {F.~X.}\ \bibnamefont {{Timmes}}}, \ and\ \bibinfo {author}
  {\bibfnamefont {R.}~\bibnamefont {{Townsend}}},\ }\href {\doibase
  10.1088/0067-0049/208/1/4} {\bibfield  {journal} {\bibinfo  {journal}
  {\apjs}\ }\textbf {\bibinfo {volume} {208}},\ \bibinfo {eid} {4} (\bibinfo
  {year} {2013})},\ \Eprint {http://arxiv.org/abs/1301.0319} {arXiv:1301.0319
  [astro-ph.SR]} \BibitemShut {NoStop}%
\bibitem [{\citenamefont {{Paxton}}\ \emph {et~al.}(2015)\citenamefont
  {{Paxton}}, \citenamefont {{Marchant}}, \citenamefont {{Schwab}},
  \citenamefont {{Bauer}}, \citenamefont {{Bildsten}}, \citenamefont
  {{Cantiello}}, \citenamefont {{Dessart}}, \citenamefont {{Farmer}},
  \citenamefont {{Hu}}, \citenamefont {{Langer}}, \citenamefont {{Townsend}},
  \citenamefont {{Townsley}},\ and\ \citenamefont
  {{Timmes}}}]{2015ApJS..220...15P}%
  \BibitemOpen
  \bibfield  {author} {\bibinfo {author} {\bibfnamefont {B.}~\bibnamefont
  {{Paxton}}}, \bibinfo {author} {\bibfnamefont {P.}~\bibnamefont
  {{Marchant}}}, \bibinfo {author} {\bibfnamefont {J.}~\bibnamefont
  {{Schwab}}}, \bibinfo {author} {\bibfnamefont {E.~B.}\ \bibnamefont
  {{Bauer}}}, \bibinfo {author} {\bibfnamefont {L.}~\bibnamefont {{Bildsten}}},
  \bibinfo {author} {\bibfnamefont {M.}~\bibnamefont {{Cantiello}}}, \bibinfo
  {author} {\bibfnamefont {L.}~\bibnamefont {{Dessart}}}, \bibinfo {author}
  {\bibfnamefont {R.}~\bibnamefont {{Farmer}}}, \bibinfo {author}
  {\bibfnamefont {H.}~\bibnamefont {{Hu}}}, \bibinfo {author} {\bibfnamefont
  {N.}~\bibnamefont {{Langer}}}, \bibinfo {author} {\bibfnamefont {R.~H.~D.}\
  \bibnamefont {{Townsend}}}, \bibinfo {author} {\bibfnamefont {D.~M.}\
  \bibnamefont {{Townsley}}}, \ and\ \bibinfo {author} {\bibfnamefont {F.~X.}\
  \bibnamefont {{Timmes}}},\ }\href {\doibase 10.1088/0067-0049/220/1/15}
  {\bibfield  {journal} {\bibinfo  {journal} {\apjs}\ }\textbf {\bibinfo
  {volume} {220}},\ \bibinfo {eid} {15} (\bibinfo {year} {2015})},\ \Eprint
  {http://arxiv.org/abs/1506.03146} {arXiv:1506.03146 [astro-ph.SR]}
  \BibitemShut {NoStop}%
\bibitem [{\citenamefont {{Paxton}}\ \emph {et~al.}(2018)\citenamefont
  {{Paxton}}, \citenamefont {{Schwab}}, \citenamefont {{Bauer}}, \citenamefont
  {{Bildsten}}, \citenamefont {{Blinnikov}}, \citenamefont {{Duffell}},
  \citenamefont {{Farmer}}, \citenamefont {{Goldberg}}, \citenamefont
  {{Marchant}}, \citenamefont {{Sorokina}}, \citenamefont {{Thoul}},
  \citenamefont {{Townsend}},\ and\ \citenamefont
  {{Timmes}}}]{2018ApJS..234...34P}%
  \BibitemOpen
  \bibfield  {author} {\bibinfo {author} {\bibfnamefont {B.}~\bibnamefont
  {{Paxton}}}, \bibinfo {author} {\bibfnamefont {J.}~\bibnamefont {{Schwab}}},
  \bibinfo {author} {\bibfnamefont {E.~B.}\ \bibnamefont {{Bauer}}}, \bibinfo
  {author} {\bibfnamefont {L.}~\bibnamefont {{Bildsten}}}, \bibinfo {author}
  {\bibfnamefont {S.}~\bibnamefont {{Blinnikov}}}, \bibinfo {author}
  {\bibfnamefont {P.}~\bibnamefont {{Duffell}}}, \bibinfo {author}
  {\bibfnamefont {R.}~\bibnamefont {{Farmer}}}, \bibinfo {author}
  {\bibfnamefont {J.~A.}\ \bibnamefont {{Goldberg}}}, \bibinfo {author}
  {\bibfnamefont {P.}~\bibnamefont {{Marchant}}}, \bibinfo {author}
  {\bibfnamefont {E.}~\bibnamefont {{Sorokina}}}, \bibinfo {author}
  {\bibfnamefont {A.}~\bibnamefont {{Thoul}}}, \bibinfo {author} {\bibfnamefont
  {R.~H.~D.}\ \bibnamefont {{Townsend}}}, \ and\ \bibinfo {author}
  {\bibfnamefont {F.~X.}\ \bibnamefont {{Timmes}}},\ }\href {\doibase
  10.3847/1538-4365/aaa5a8} {\bibfield  {journal} {\bibinfo  {journal} {\apjs}\
  }\textbf {\bibinfo {volume} {234}},\ \bibinfo {eid} {34} (\bibinfo {year}
  {2018})},\ \Eprint {http://arxiv.org/abs/1710.08424} {arXiv:1710.08424
  [astro-ph.SR]} \BibitemShut {NoStop}%
\bibitem [{\citenamefont {{Paxton}}\ \emph {et~al.}(2019)\citenamefont
  {{Paxton}}, \citenamefont {{Smolec}}, \citenamefont {{Schwab}}, \citenamefont
  {{Gautschy}}, \citenamefont {{Bildsten}}, \citenamefont {{Cantiello}},
  \citenamefont {{Dotter}}, \citenamefont {{Farmer}}, \citenamefont
  {{Goldberg}}, \citenamefont {{Jermyn}}, \citenamefont {{Kanbur}},
  \citenamefont {{Marchant}}, \citenamefont {{Thoul}}, \citenamefont
  {{Townsend}}, \citenamefont {{Wolf}}, \citenamefont {{Zhang}},\ and\
  \citenamefont {{Timmes}}}]{2019ApJS..243...10P}%
  \BibitemOpen
  \bibfield  {author} {\bibinfo {author} {\bibfnamefont {B.}~\bibnamefont
  {{Paxton}}}, \bibinfo {author} {\bibfnamefont {R.}~\bibnamefont {{Smolec}}},
  \bibinfo {author} {\bibfnamefont {J.}~\bibnamefont {{Schwab}}}, \bibinfo
  {author} {\bibfnamefont {A.}~\bibnamefont {{Gautschy}}}, \bibinfo {author}
  {\bibfnamefont {L.}~\bibnamefont {{Bildsten}}}, \bibinfo {author}
  {\bibfnamefont {M.}~\bibnamefont {{Cantiello}}}, \bibinfo {author}
  {\bibfnamefont {A.}~\bibnamefont {{Dotter}}}, \bibinfo {author}
  {\bibfnamefont {R.}~\bibnamefont {{Farmer}}}, \bibinfo {author}
  {\bibfnamefont {J.~A.}\ \bibnamefont {{Goldberg}}}, \bibinfo {author}
  {\bibfnamefont {A.~S.}\ \bibnamefont {{Jermyn}}}, \bibinfo {author}
  {\bibfnamefont {S.~M.}\ \bibnamefont {{Kanbur}}}, \bibinfo {author}
  {\bibfnamefont {P.}~\bibnamefont {{Marchant}}}, \bibinfo {author}
  {\bibfnamefont {A.}~\bibnamefont {{Thoul}}}, \bibinfo {author} {\bibfnamefont
  {R.~H.~D.}\ \bibnamefont {{Townsend}}}, \bibinfo {author} {\bibfnamefont
  {W.~M.}\ \bibnamefont {{Wolf}}}, \bibinfo {author} {\bibfnamefont
  {M.}~\bibnamefont {{Zhang}}}, \ and\ \bibinfo {author} {\bibfnamefont
  {F.~X.}\ \bibnamefont {{Timmes}}},\ }\href {\doibase
  10.3847/1538-4365/ab2241} {\bibfield  {journal} {\bibinfo  {journal} {\apjs}\
  }\textbf {\bibinfo {volume} {243}},\ \bibinfo {eid} {10} (\bibinfo {year}
  {2019})},\ \Eprint {http://arxiv.org/abs/1903.01426} {arXiv:1903.01426
  [astro-ph.SR]} \BibitemShut {NoStop}%
\bibitem [{\citenamefont {{Jermyn}}\ \emph {et~al.}(2023)\citenamefont
  {{Jermyn}}, \citenamefont {{Bauer}}, \citenamefont {{Schwab}}, \citenamefont
  {{Farmer}}, \citenamefont {{Ball}}, \citenamefont {{Bellinger}},
  \citenamefont {{Dotter}}, \citenamefont {{Joyce}}, \citenamefont
  {{Marchant}}, \citenamefont {{Mombarg}}, \citenamefont {{Wolf}},
  \citenamefont {{Sunny Wong}}, \citenamefont {{Cinquegrana}}, \citenamefont
  {{Farrell}}, \citenamefont {{Smolec}}, \citenamefont {{Thoul}}, \citenamefont
  {{Cantiello}}, \citenamefont {{Herwig}}, \citenamefont {{Toloza}},
  \citenamefont {{Bildsten}}, \citenamefont {{Townsend}},\ and\ \citenamefont
  {{Timmes}}}]{2023ApJS..265...15J}%
  \BibitemOpen
  \bibfield  {author} {\bibinfo {author} {\bibfnamefont {A.~S.}\ \bibnamefont
  {{Jermyn}}}, \bibinfo {author} {\bibfnamefont {E.~B.}\ \bibnamefont
  {{Bauer}}}, \bibinfo {author} {\bibfnamefont {J.}~\bibnamefont {{Schwab}}},
  \bibinfo {author} {\bibfnamefont {R.}~\bibnamefont {{Farmer}}}, \bibinfo
  {author} {\bibfnamefont {W.~H.}\ \bibnamefont {{Ball}}}, \bibinfo {author}
  {\bibfnamefont {E.~P.}\ \bibnamefont {{Bellinger}}}, \bibinfo {author}
  {\bibfnamefont {A.}~\bibnamefont {{Dotter}}}, \bibinfo {author}
  {\bibfnamefont {M.}~\bibnamefont {{Joyce}}}, \bibinfo {author} {\bibfnamefont
  {P.}~\bibnamefont {{Marchant}}}, \bibinfo {author} {\bibfnamefont {J.~S.~G.}\
  \bibnamefont {{Mombarg}}}, \bibinfo {author} {\bibfnamefont {W.~M.}\
  \bibnamefont {{Wolf}}}, \bibinfo {author} {\bibfnamefont {T.~L.}\
  \bibnamefont {{Sunny Wong}}}, \bibinfo {author} {\bibfnamefont {G.~C.}\
  \bibnamefont {{Cinquegrana}}}, \bibinfo {author} {\bibfnamefont
  {E.}~\bibnamefont {{Farrell}}}, \bibinfo {author} {\bibfnamefont
  {R.}~\bibnamefont {{Smolec}}}, \bibinfo {author} {\bibfnamefont
  {A.}~\bibnamefont {{Thoul}}}, \bibinfo {author} {\bibfnamefont
  {M.}~\bibnamefont {{Cantiello}}}, \bibinfo {author} {\bibfnamefont
  {F.}~\bibnamefont {{Herwig}}}, \bibinfo {author} {\bibfnamefont
  {O.}~\bibnamefont {{Toloza}}}, \bibinfo {author} {\bibfnamefont
  {L.}~\bibnamefont {{Bildsten}}}, \bibinfo {author} {\bibfnamefont {R.~H.~D.}\
  \bibnamefont {{Townsend}}}, \ and\ \bibinfo {author} {\bibfnamefont {F.~X.}\
  \bibnamefont {{Timmes}}},\ }\href {\doibase 10.3847/1538-4365/acae8d}
  {\bibfield  {journal} {\bibinfo  {journal} {\apjs}\ }\textbf {\bibinfo
  {volume} {265}},\ \bibinfo {eid} {15} (\bibinfo {year} {2023})},\ \Eprint
  {http://arxiv.org/abs/2208.03651} {arXiv:2208.03651 [astro-ph.SR]}
  \BibitemShut {NoStop}%
\bibitem [{\citenamefont {{Serenelli}}\ \emph {et~al.}(2017)\citenamefont
  {{Serenelli}}, \citenamefont {{Weiss}}, \citenamefont {{Cassisi}},
  \citenamefont {{Salaris}},\ and\ \citenamefont
  {{Pietrinferni}}}]{Serenelli:2017}%
  \BibitemOpen
  \bibfield  {author} {\bibinfo {author} {\bibfnamefont {A.}~\bibnamefont
  {{Serenelli}}}, \bibinfo {author} {\bibfnamefont {A.}~\bibnamefont
  {{Weiss}}}, \bibinfo {author} {\bibfnamefont {S.}~\bibnamefont {{Cassisi}}},
  \bibinfo {author} {\bibfnamefont {M.}~\bibnamefont {{Salaris}}}, \ and\
  \bibinfo {author} {\bibfnamefont {A.}~\bibnamefont {{Pietrinferni}}},\ }\href
  {\doibase 10.1051/0004-6361/201731004} {\bibfield  {journal} {\bibinfo
  {journal} {\aap}\ }\textbf {\bibinfo {volume} {606}},\ \bibinfo {eid} {A33}
  (\bibinfo {year} {2017})},\ \Eprint {http://arxiv.org/abs/1706.09910}
  {arXiv:1706.09910 [astro-ph.SR]} \BibitemShut {NoStop}%
\bibitem [{\citenamefont {Fung}\ \emph {et~al.}(2024)\citenamefont {Fung},
  \citenamefont {Heeba}, \citenamefont {Liu}, \citenamefont {Muralidharan},
  \citenamefont {Schutz},\ and\ \citenamefont {Vincent}}]{Fung:2023euv}%
  \BibitemOpen
  \bibfield  {author} {\bibinfo {author} {\bibfnamefont {A.}~\bibnamefont
  {Fung}}, \bibinfo {author} {\bibfnamefont {S.}~\bibnamefont {Heeba}},
  \bibinfo {author} {\bibfnamefont {Q.}~\bibnamefont {Liu}}, \bibinfo {author}
  {\bibfnamefont {V.}~\bibnamefont {Muralidharan}}, \bibinfo {author}
  {\bibfnamefont {K.}~\bibnamefont {Schutz}}, \ and\ \bibinfo {author}
  {\bibfnamefont {A.~C.}\ \bibnamefont {Vincent}},\ }\href {\doibase
  10.1103/PhysRevD.109.083011} {\bibfield  {journal} {\bibinfo  {journal}
  {Phys. Rev. D}\ }\textbf {\bibinfo {volume} {109}},\ \bibinfo {pages}
  {083011} (\bibinfo {year} {2024})},\ \Eprint
  {http://arxiv.org/abs/2309.06465} {arXiv:2309.06465 [hep-ph]} \BibitemShut
  {NoStop}%
\bibitem [{\citenamefont {Huang}\ \emph {et~al.}(2023)\citenamefont {Huang}
  \emph {et~al.}}]{PandaX:2023xgl}%
  \BibitemOpen
  \bibfield  {author} {\bibinfo {author} {\bibfnamefont {D.}~\bibnamefont
  {Huang}} \emph {et~al.} (\bibinfo {collaboration} {PandaX}),\ }\href
  {\doibase 10.1103/PhysRevLett.131.191002} {\bibfield  {journal} {\bibinfo
  {journal} {Phys. Rev. Lett.}\ }\textbf {\bibinfo {volume} {131}},\ \bibinfo
  {pages} {191002} (\bibinfo {year} {2023})},\ \Eprint
  {http://arxiv.org/abs/2308.01540} {arXiv:2308.01540 [hep-ex]} \BibitemShut
  {NoStop}%
\bibitem [{\citenamefont {Erickcek}\ \emph {et~al.}(2007)\citenamefont
  {Erickcek}, \citenamefont {Steinhardt}, \citenamefont {McCammon},\ and\
  \citenamefont {McGuire}}]{Erickcek:2007jv}%
  \BibitemOpen
  \bibfield  {author} {\bibinfo {author} {\bibfnamefont {A.~L.}\ \bibnamefont
  {Erickcek}}, \bibinfo {author} {\bibfnamefont {P.~J.}\ \bibnamefont
  {Steinhardt}}, \bibinfo {author} {\bibfnamefont {D.}~\bibnamefont
  {McCammon}}, \ and\ \bibinfo {author} {\bibfnamefont {P.~C.}\ \bibnamefont
  {McGuire}},\ }\href {\doibase 10.1103/PhysRevD.76.042007} {\bibfield
  {journal} {\bibinfo  {journal} {Phys. Rev. D}\ }\textbf {\bibinfo {volume}
  {76}},\ \bibinfo {pages} {042007} (\bibinfo {year} {2007})},\ \Eprint
  {http://arxiv.org/abs/0704.0794} {arXiv:0704.0794 [astro-ph]} \BibitemShut
  {NoStop}%
\bibitem [{\citenamefont {Emken}\ and\ \citenamefont
  {Kouvaris}(2018)}]{Emken:2018run}%
  \BibitemOpen
  \bibfield  {author} {\bibinfo {author} {\bibfnamefont {T.}~\bibnamefont
  {Emken}}\ and\ \bibinfo {author} {\bibfnamefont {C.}~\bibnamefont
  {Kouvaris}},\ }\href {\doibase 10.1103/PhysRevD.97.115047} {\bibfield
  {journal} {\bibinfo  {journal} {Phys. Rev. D}\ }\textbf {\bibinfo {volume}
  {97}},\ \bibinfo {pages} {115047} (\bibinfo {year} {2018})},\ \Eprint
  {http://arxiv.org/abs/1802.04764} {arXiv:1802.04764 [hep-ph]} \BibitemShut
  {NoStop}%
\end{thebibliography}%
\end{document}